\documentclass[12pt,aps,prd,superscriptaddress,floatfix,showpacs,preprint,nofootinbib]{revtex4}

\usepackage[dvips]{graphicx}
\usepackage{amsmath}
\usepackage{slashed}
\allowdisplaybreaks
\newcommand{\fig}[1]{\mbox{\textcircled{{\scriptsize #1}}}}

\begin{document}
\bibliographystyle{apsrev}
\preprint{CU-TP-1197, MPP-2011-51}

\title{$K$ to $\pi\pi$ Decay amplitudes from Lattice QCD}

\newcommand\bnl{Brookhaven National Laboratory, Upton, NY 11973, USA}
\newcommand\cu{Physics Department, Columbia University, New York,
      NY 10027, USA}
\newcommand\edinb{SUPA, School of Physics, The University of
  Edinburgh, Edinburgh EH9 3JZ, UK}
\newcommand{\indiana}{Department of Physics, Indiana University,
      Bloomington, IN 47405, USA}
\newcommand{\mpimunchen}{Max-Planck-Institut f{\"u}r Physik, 
                 F{\"o}hringer Ring 6, 80805 M{\"u}nchen, Germany}
\newcommand\riken{RIKEN-BNL Research Center, Brookhaven National
      Laboratory, Upton, NY 11973, USA}
\newcommand\soton{School of Physics and Astronomy, University of
  Southampton,  Southampton SO17 1BJ, UK}
\newcommand\uconn{Physics Department, University of Connecticut,
      Storrs, CT 06269-3046, USA}

\author{T.~Blum}\affiliation{\uconn}
\author{P.A.~Boyle}\affiliation{\edinb}
\author{N.H.~Christ}\affiliation{\cu}
\author{N.~Garron}\affiliation{\edinb}
\author{E.~Goode}\affiliation{\soton}
\author{T.~Izubuchi}\affiliation{\bnl}\affiliation{\riken}
\author{C.~Lehner}\affiliation{\riken}
\author{Q.~Liu}\affiliation{\cu}
\author{R.D.~Mawhinney}\affiliation{\cu}
\author{C.T.~Sachrajda}\affiliation{\soton}
\author{A.~Soni}\affiliation{\bnl}
\author{C.~Sturm}\affiliation{\mpimunchen}
\author{H.~Yin}\affiliation{\cu}
\author{R.~Zhou}\affiliation{\uconn}\affiliation{\indiana}
\collaboration{RBC and UKQCD Collaborations}
%Chris I added "Collaborations".

\date{June 09, 2011}

\begin{abstract}
We report a direct lattice calculation of the $K$ to $\pi\pi$ decay
matrix elements for both the $\Delta I=1/2$ and $3/2$ amplitudes $A_0$
and $A_2$ on 2+1 flavor, domain wall fermion, $16^3\times32\times16$
lattices.  This is a complete calculation in which all contractions
for the required ten, four-quark operators are evaluated, including
the disconnected graphs in which no quark line connects the initial
kaon and final two-pion states.  These lattice operators are
non-perturbatively renormalized using the Rome-Southampton method
and the quadratic divergences are studied and removed.  This is an
important but notoriously difficult calculation, requiring high
statistics on a large volume.  In this paper we take a major step 
towards the computation of the physical $K\to\pi\pi$ amplitudes by 
performing a complete calculation at unphysical
kinematics with pions of mass 422\,MeV at rest in the kaon rest
frame.  With this simplification we are able to resolve Re$(A_0)$
from zero for the first time, with a 25\% statistical error and can
develop and evaluate methods for computing the complete, complex
amplitude $A_0$, a calculation central to understanding the
$\Delta =1/2$ rule and testing the standard model of CP violation
in the kaon system.
\end{abstract}

\pacs{11.15.Ha, % Lattice gauge theory
%     11.30.Rd, % Chiral symmetries
      12.38.Gc  % Lattice QCD calculations
      14.40.Be  % Light mesons (S=C=B=0)
      13.25.Es  % Hadronic kaon decays
}

\maketitle

\newpage

\section{Introduction}

The Cabibbo-Kobayashi-Maskawa (CKM) theory for the weak interactions of the
quarks when combined with QCD provides a framework describing in complete
detail all the properties and interactions of the six quarks.  This framework
incorporates the most general assignment of masses and couplings and appears
able to explain all observed phenomena in which these quarks participate.
However, to date, the non-perturbative character of low energy QCD has
obscured many of the consequences of the CKM theory.  In particular, both the
direct CP violation seen in K meson decay and the factor of 22.5 enhancement
of the $I=0$, $K\rightarrow\pi\pi$ decay amplitude $A_0$ relative to the
$I=2$ amplitude $A_2$ (the $\Delta I=1/2$ rule) lack a quantitative explanation.

Wilson coefficients evaluated at a QCD scale of about $2$ GeV represent
the short distance physics and can be evaluated from the CKM theory using
QCD and electro-weak perturbation theory.  However, these factors explain
only a factor of two enhancement of the $I=0$ amplitude~\cite{Gaillard:1974nj,
Altarelli:1974exa}.  The remaining enhancement must arise from the hadronic
matrix elements which require non-perturbative treatment.

Direct CP violation in kaon decays provides
a critical test of the standard model's CKM mechanism of CP violation.
While forty years of experimental effort have produced the measured result
Re$(\epsilon'/\epsilon)=1.65(26)\times10^{-3}$~\cite{Nakamura:2010zzi}, with
only a 16\% error, there is no reliable theoretical calculation of this quantity
based on the standard model.  A previous lattice QCD calculation using
2+1 dynamical domain wall fermions failed to give a conclusive result
because of the large systematic errors associated with the use of chiral
perturbation theory at the scale of the kaon mass~\cite{Li:2008kc}.
(However, there are on-going efforts using chiral perturbation 
theory~\cite{Laiho:2010ir}.)  Earlier quenched results~\cite{Blum:2001xb,Noaki:2001un}
are subject to this same difficulty together with uncontrolled uncertainties
associated with quenching~\cite{Golterman:2001qj,Golterman:2002us,Aubin:2006vt}.

A direct lattice calculation of $K\rightarrow\pi\pi$ decay is
extremely important to provide an explanation for the $\Delta I=1/2$ rule
and to test the standard model of CP violation from first principles.
This is an unusually difficult calculation because of the presence of
disconnected graphs.  However, with the continuing increase of available
computing power and the development of improved algorithms, calculations
with disconnected graphs are now no longer out of reach.  In fact, our
recent successful calculation of the masses and mixing of the $\eta^\prime$
and $\eta$ mesons~\cite{Christ:2010dd} was carried out in part to develop
and test the methods needed for the calculation presented here.
In this paper, we present a first direct calculation of the complete
$K^0 \rightarrow \pi\pi$ decay amplitude. At this stage, we work with the simplified kinematics of a threshold decay in which the kaon is at rest 
and decays into two pions each with zero momentum and with mass one-half 
that of the kaon. The calculation with this choice of kinematics still 
contains the main difficulties we need to overcome in order to be able 
to compute the physical $K\to\pi\pi$ decay amplitudes; i.e. the presence 
of disconnected diagrams coupled with the need to subtract ultraviolet 
power divergences. However, as explained below, with the pions at rest 
we are able to generate sufficient statistics to explore how to handle 
these difficulties. We stress that at this simplified choice of kinematics, 
we compute the $K\to\pi\pi$ amplitudes directly and completely.

In order to calculate the decay amplitudes, we perform a direct, brute force
calculation of the required weak matrix elements. The isospin zero $\pi-\pi$
final state implies the presence of disconnected graphs in correlation 
functions and makes the calculation very difficult.   For these graphs, 
the noise does not decrease with increasing time separation between the 
source and sink, while the signal does.  Therefore, substantial statistics 
are needed to get a clear signal.  This difficulty is compounded by the 
presence of diagrams which diverge as $1/a^2$ as the continuum limit is 
approached ($a$ is the lattice spacing).  While these divergent amplitudes 
must vanish for a physical, on-shell decay they substantially degrade the 
signal to noise ratio even for an energy-conserving calculation such as 
this one.  Studying the properties of the $1/a^2$ terms and learning how 
to successfully subtract them is one of the important objectives of this 
calculation.  The chiral symmetry needed to control operator mixing is 
provided by our use of domain wall fermions.

Recognizing the difficulty of this problem, we choose to perform this first
calculation on a lattice which is relatively small compared to those used in
other recent work and to use a somewhat heavy pion mass ($m_\pi\approx$
421\,MeV) so we can more easily collect large statistics.  We concentrate
on exploring and reducing the statistical uncertainty since the primary
goal of this work is to extract a clear signal for these amplitudes.
Therefore, the quoted errors on our results are statistical only.

The main objective of this paper is to calculate the $\Delta I=1/2$ decay
amplitude $A_0$.  A calculation of the $\Delta I=3/2$ part is included here
for comparison and completeness.  A much more physical calculation of this
$\Delta I=3/2$ amplitude alone can be found in~\cite{Goode:2011kb}.  In
the case of the $I=2$ final state no disconnected diagrams appear, there
are no divergent eye diagrams and isospin conservation requires that four
valence quark propagators must join the kaon and weak operator with the
operators creating the two final-state pions.  This allows physical
kinematics with non-zero final momenta to be achieved by imposing
anti-periodic boundary conditions on one species of valence
quark~\cite{Kim:2003xt,Sachrajda:2004mi}.  As a result, the preliminary
calculation of $A_2$ reported in Ref.~\cite{Goode:2011kb} is performed
at almost physical kinematics on a lattice of spatial size 4.5 fm
and determines complex $A_2$ with controlled errors of $O(10\%)$.  The
present work is intended as the first step toward an equally physical
but much more challenging calculation of $A_0$.

While we do not employ physical kinematics, the final results for the
complex amplitudes $A_0$ and $A_2$ presented in this paper are otherwise 
physical.   In particular, we use Rome-Southampton methods~\cite{Martinelli:1995ty}
to change the normalization of our bare lattice four-quark operators to
that of the RI/MOM scheme.  A second conversion to the $\overline{\mbox{MS}}$ 
scheme is then performed using the recent results of Ref.~\cite{Lehner:2011fz}.  
Finally these $\overline{\mbox{MS}}$-normalized matrix elements are combined 
with the appropriate Wilson coefficients~\cite{Buchalla:1995vs}, determined 
in this same scheme, to obtain our results for $A_0$ and $A_2$.  Because of 
our unphysical, threshold kinematics and focus on controlling the statistical 
errors associated with the disconnected diagrams, we do not estimate the size
of possible systematic errors.  Similarly we do not include the systematic 
or statistical errors associated with the Rome-Southampton renormalization
factors, both of which could be made substantially than our statistical 
errors when required.

This paper is organized as follows.  We first summarize our computational
setup, including our strategy to collect large statistics.  Next we discuss
our results for $\pi-\pi$ scattering which are a by-product of the necessary
characterization of the operator creating the $\pi-\pi$ final state and
are also needed to evaluate the Lellouch-L\"uscher, finite-volume correction
\cite{Lellouch:2000pv}.  After a section giving the details of the
$K^0 \rightarrow \pi\pi$ contractions, we provide our numerical results for the
$K^0 \rightarrow \pi\pi$ decay amplitudes for both the $\Delta I=3/2$ and
$1/2$ channels.  The details of the operator renormalization required by
the Wilson coefficients which we use are presented in Appendix A.  Finally 
we present our conclusions and discuss future prospects.

\section{Computational Details}
\label{sec:Comp_details}

Our calculation uses the Iwasaki gauge action with $\beta=2.13$ and 2+1 flavors
of domain wall fermions (DWF).  While the computational costs of DWF are much
greater than those of Wilson or staggered fermions, as has been shown in
earlier papers~\cite{Bernard:1988zj,Dawson:1997ic,Blum:2001xb,Noaki:2001un},
accurate chiral symmetry at short distances is critical to avoid extensive
operator mixing, which would make the lattice treatment of $\Delta S = 1$
processes much more difficult.

We use a single lattice ensemble with space-time volume $16^3\times32$, a
fifth-dimensional extent of $L_s=16$ and light and strange quark masses of
$m_l=0.01$, $m_s=0.032$, respectively.  This ensemble is similar to the
$m_l=0.01$ ensemble reported in Ref.~\cite{Allton:2007hx} except we use
the improved RHMC-II algorithm of Ref.~\cite{Allton:2008pn} and a more
physical value for the strange quark mass.  The inverse lattice spacing
for these input parameters was determined to be 1.73(3)GeV and the
residual mass is $m_{\rm res}=0.00308(4)$~\cite{Allton:2008pn}.  The total
number of configurations we used is 800, each separated by 10 time units.
We initially generated an ensemble one-half of this size.  When our analysis
showed a non-zero result for Re$A_0$, we then doubled the size of the
ensemble to assure ourselves that the result was trustworthy and to reduce
the resulting error.  We have performed the analysis described below both
by treating the results from each configuration as independent and by
grouping them into blocks.  The resulting statistical errors are independent
of block size suggesting that the individual configurations are essentially
uncorrelated for our observables.

We use anti-periodic boundary conditions in the time direction, and
periodic boundary conditions in the space directions for the Dirac
operator.  The propagators (inverses of the Dirac operator) are
calculated using a Coulomb gauge fixed wall source (used for meson
propagators) and a random wall source (used to calculate the loops
in the $type3$ and $type4$ graphs shown in Figs.~\ref{fig:type3} and
\ref{fig:type4} below) for each of the 32 time slices in our lattice volume.
For each time slice and source type, twelve inversions are required
corresponding to the possible 3 color and 4 spin choices for the
source.  Thus, all together we carry out 768 inversions for each quark
mass on a given configuration.  As will be shown below, this large number 
of inversions, performed on 800 configurations, provides the substantial
statistics needed to resolve the real part of the $I=0$ amplitude
$A_0$ with $25\%$ accuracy.

The situation described above in which 768 Dirac propagators must
be computed on a single gauge background is an excellent candidate
for the use of deflation techniques.  The overhead associated with
determining a set of low eigenmodes of this single Dirac operator
can be effectively amortized over the many inversions in which those
low modes can be used.  Our $m_l=0.01$, light quark inversions are
accelerated by a factor of 2-3 by using exact, low-mode deflation~\cite{Giusti:2002sm} 
in which we compute the Dirac eigenvectors with the smallest 35 eigenvalues
and limit the conjugate gradient inversion to the remaining orthogonal
subspace.

\begin{table*}
\caption{Masses of pion and kaons and energies of the two-pion states.
Here the subscript $I=0$ or 2 on the $\pi-\pi$ energy, $E_I^{\pi\pi}$,
labels the isospin of the state and $E_0^{\pi\pi\prime}$ represents the
isospin zero, two-pion energy obtained when the disconnected graph V is
ignored.  The superscript (0), (1) or (2) on the kaon mass distinguishes our three choices of valence strange quark mass, $m_s = 0.066$, 0.099 and 
0.165 respectively.
}
\label{tab:mass}
\begin{ruledtabular}
  \begin{tabular}{lllllll}
$m_\pi$     & $E_0^{\pi\pi}$
                        & $E_0^{\pi\pi\prime}$
                                    & $E_2^{\pi\pi}$
                                                 & $m_K^{(0)}$ & $m_K^{(1)}$ & $m_K^{(2)}$ \\
\hline
0.24373(47) & 0.443(13) & 0.4393(41)& 0.5066(11) & 0.42599(42) & 0.50729(44) & 0.64540(49) \\
\end{tabular}
\end{ruledtabular}
\end{table*}

In order to obtain energy-conserving $K^0\rightarrow\pi\pi$ decay
amplitudes, the mass of the valence strange quark in the kaon is
assigned a value different from that appearing in the fermion determinant
used to generate the ensembles, {\it i.e.} the strange quark is partially
quenched.  Since the mass of the dynamical strange quark is expected to
have a small effect on amplitudes of the sort considered here~\cite{Allton:2008pn,
Lightman:2009ka}, this use of partial quenching is appropriate for the
purposes of this paper.  Valence strange quark masses are chosen to be
$m_s=0.066$, 0.099 and 0.165, which are labeled 0, 1 and 2 respectively.
The resulting kaon masses are shown in Tab.~\ref{tab:mass}. In the
following section we will see that by using these values for $m_s$ we can
interpolate to energy-conserving decay kinematics for both the $I=2$
and $I=0$ channels.

\section{Two-pion Scattering}

\begin{figure}
\begin{minipage}[c]{0.40\textwidth}
\includegraphics[width=0.8\textwidth]{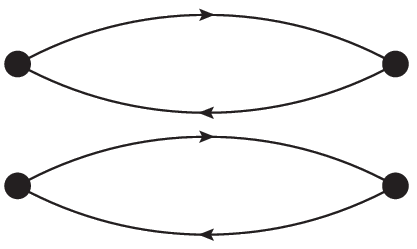}
\end{minipage}
\begin{minipage}[c]{0.40\textwidth}
\includegraphics[width=0.8\textwidth]{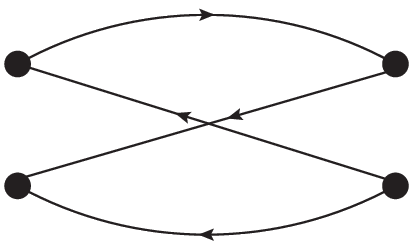}
\end{minipage} \\
\begin{minipage}[c]{0.40\textwidth}
\includegraphics[width=0.8\textwidth]{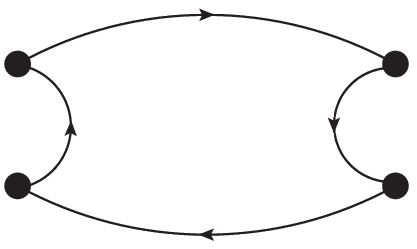}
\end{minipage}
\begin{minipage}[c]{0.40\textwidth}
\includegraphics[width=0.8\textwidth]{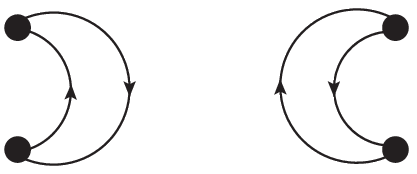}
\end{minipage}
\caption{The four diagrams which contribute to $\pi-\pi$ scattering:
direct (D), cross (C), rectangle (R), and vacuum (V), arranged from the
left top to right bottom.}
\label{fig:DCRV}
\end{figure}

The $\pi-\pi$ scattering calculation requires 4 contractions which we
have labeled direct (D), cross (C), rectangle (R), and vacuum (V) as in
Ref.~\cite{Liu:2009uw} and which are shown in Fig.~\ref{fig:DCRV}.  For
convenience, the minus sign arising from the number of fermion loops is
not included in the definition of these contractions.  The vacuum
contraction should be accompanied by a vacuum subtraction.  These
contractions can be calculated in terms of the light quark propagator
$L(t_\mathrm{snk}, t_\mathrm{src})$ for a Coulomb gauge fixed wall
source located at the time $t_\mathrm{src}$ and a similar wall sink
located at $t_\mathrm{snk}$.  The resulting complete vacuum amplitude,
including the vacuum subtraction, is given by
\begin{eqnarray}
V(t) &=& \frac{1}{32}\sum_{t'=0}^{31} \Biggl\{\Bigl\langle \mbox{tr}[L(t',t')L(t',t')^\dagger]
                           \mbox{tr}[L(t+t', t+t')L(t+t', t+t')^\dagger] \Bigr\rangle
\label{eq:pi-pi_vacuum} \\
     && \hskip 0.3in - \Bigl\langle \mbox{tr}[L(t',t')L(t',t')^\dagger]\Bigr\rangle
          \Bigl\langle \mbox{tr}[L(t+t',t+t')L(t+t',t+t')^\dagger]\Bigr\rangle\Biggr\},
\nonumber
\end{eqnarray}
where the indicated traces are taken over spin and color, the hermiticity
properties of the domain wall propagator have been used to eliminate factors
of $\gamma^5$ and we are explicitly combining the results from each of the
32 time slices.

\begin{figure*}
\begin{tabular}{ll}
\includegraphics[width=0.5\textwidth]{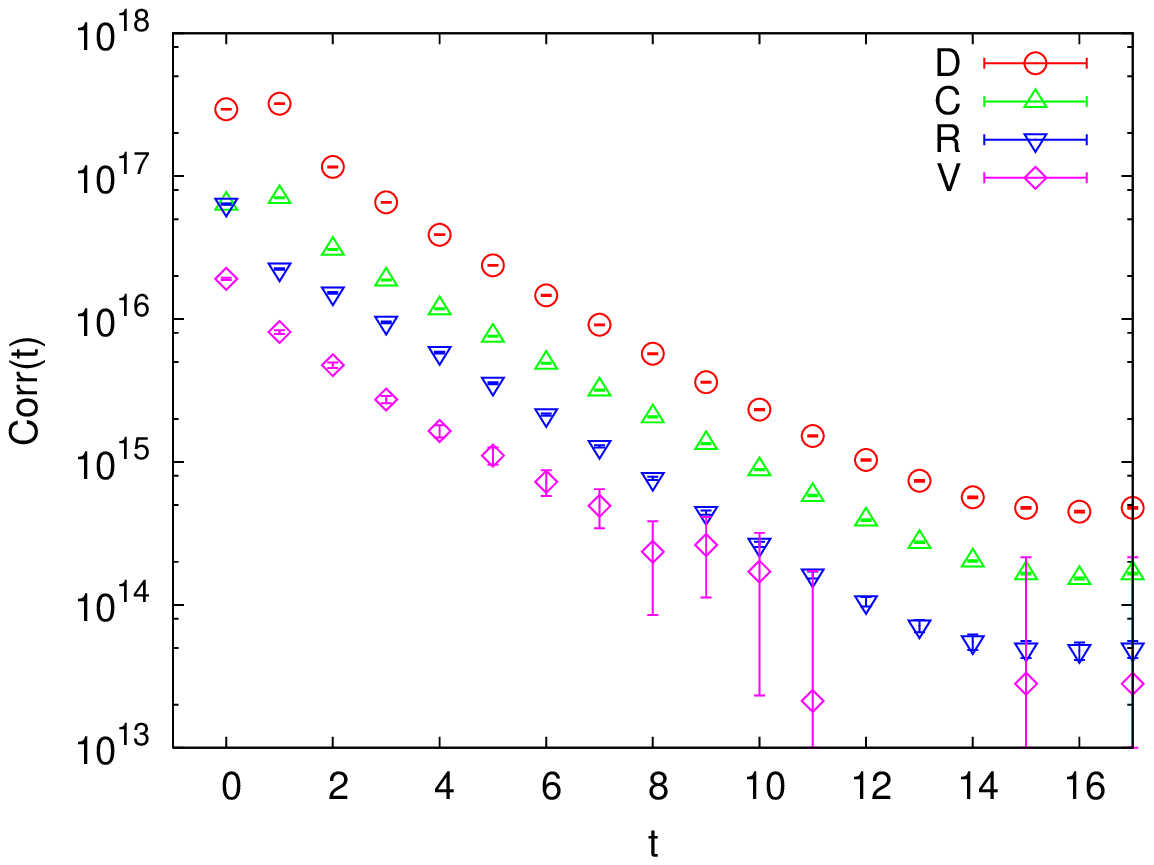} &
\includegraphics[width=0.5\textwidth]{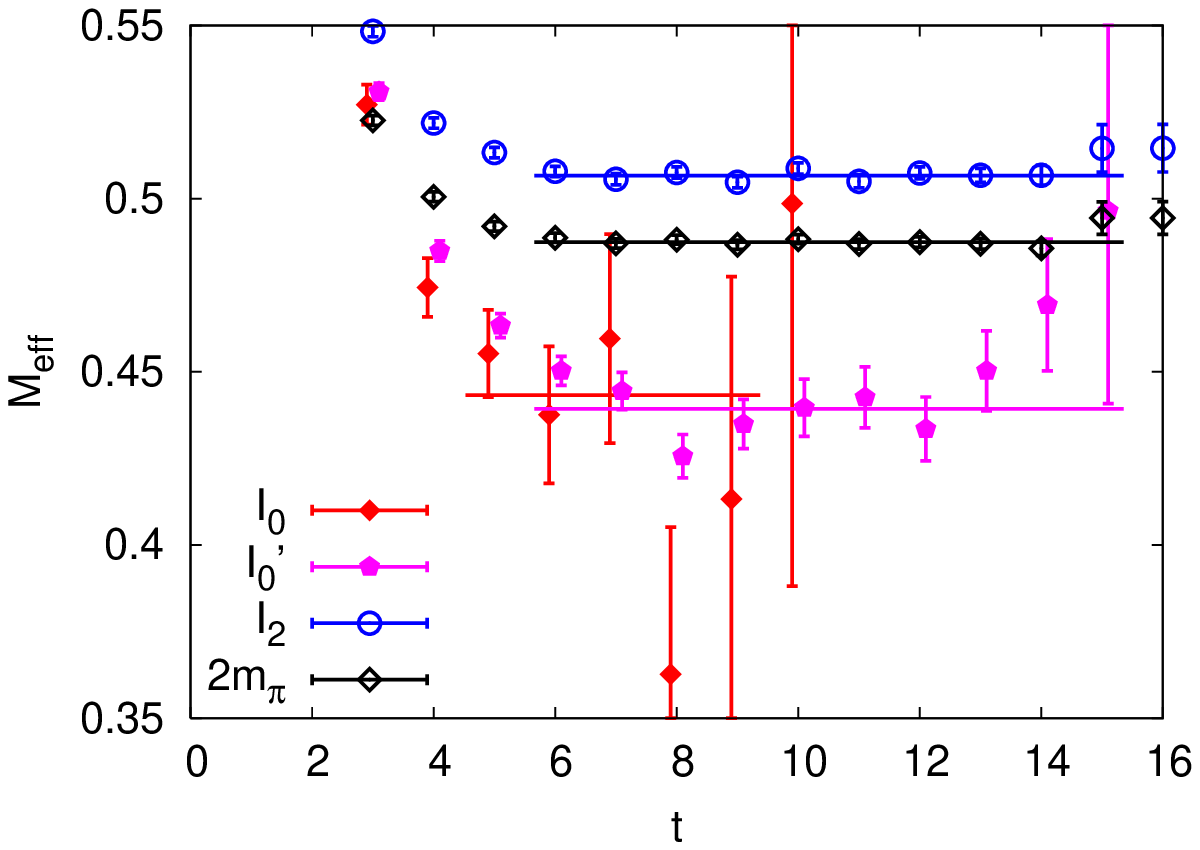}
\end{tabular}
\caption{Left: Results for the four types of contractions, direct (D),
cross (C), rectangle (R), and vacuum(V) represented by the graphs
in Fig.~\protect{\ref{fig:DCRV}}.  Right: Effective mass plots for
correlation functions for states with isospin two ($I_2$), isospin zero
($I_0$), isospin zero without the disconnected graph ($I_0^\prime$)
and twice the pion effective mass ($2m_\pi$).}
\label{fig:twopion}
\end{figure*}

Our results for each of these four types of contractions are shown in
the left panel of Fig.~\ref{fig:twopion}.  Notice that the disconnected
(vacuum) graph has an almost constant error with increasing time
separation between the source and sink, so it appears to have an
increasing error bar in the log plot, while the signal decreases
exponentially.

These four types of correlators can be combined to construct physical
correlation functions for two-pion states with definite isospin:
\begin{eqnarray}
\left\langle O_2^{\pi\pi}(t+t')^\dagger O_2^{\pi\pi}(t') \right\rangle
                & = & 2\bigl(D(t)-C(t)\bigr) \\
\left\langle O_0^{\pi\pi}(t+t')^\dagger O_0^{\pi\pi}(t') \right\rangle
                & = & 2D(t)+C(t)-6R(t)+3V(t).
\end{eqnarray}
Here the operator $O_I^{\pi\pi}(t)$ creates a two-pion state
with total isospin $I$ and $z$-component of isospin $I_z=0$ using two
quark and two anti-quark wall-sources located at the time-slice $t$.
As in Eq.~\ref{eq:pi-pi_vacuum} we will average over all 32 possible
values of common time displacement $t'$ to improve statistics.

The two-pion correlation functions for isospin $I$ and $I_z=0$ are fit
with a functional form Corr$_I(t) = N_I^2\{\exp(-E_I^{\pi\pi} t)
+\exp(-E_I^{\pi\pi}(T-t))+C_I\}$, where the constant $C_I$ comes
from the case in which the two pions propagate in opposite time
directions.  The fitted energies are summarized in Tab.~\ref{tab:mass}.
In order to see clearly the effect of the disconnected graph, we also
perform the calculation for the $I=0$ channel without the disconnected
graphs.  This result is given in Tab.~\ref{tab:mass} with a label
with an additional prime ($\prime$) symbol. The resulting effective
mass plots for each case are shown in the right panel of
Fig.~\ref{fig:twopion}.  For comparison, a plot of twice the pion
effective mass is also shown.  This figure clearly demonstrates that
the two-pion interaction is attractive in the $I=0$ channel with the
finite volume, $I=0$ $\pi-\pi$ energy $E_0^{\pi\pi}$ lower than $2m_\pi$.
In contrast, the $I=2$ channel is repulsive with $E_2^{\pi\pi}$ larger
than $2m_\pi$.  The fitted parameters $N_I^{\pi\pi}$ and $E_I^{\pi\pi}$
will be used to extract weak matrix elements from the
$K^0 \rightarrow \pi\pi$ correlation functions discussed below in
which these same operators $O_I^{\pi\pi}(t)$ are used to construct
the two-pion states.

\section{Contractions for $K^0\rightarrow\pi\pi$ Decays}
\label{sec:K_decay_contractions}

The effective weak Hamiltonian describing $K^0\rightarrow\pi\pi$ decay
including the $u$, $d$, and $s$ flavors as dynamical variables is
\begin{equation}
 H_w=\frac{G_F}{\sqrt{2}}V_{ud}^*V_{us}
                      \sum_{i=1}^{10}[(z_i(\mu)+\tau y_i(\mu))] Q_i.
\label{eq:weak_eff}
\end{equation}
Throughout this paper we follow the conventions and notation of
Ref.~\cite{Blum:2001xb}.  In Eq.~\ref{eq:weak_eff} the $Q_i$ are the
ten conventional four-quark operators, $z_i$ and $y_i$ are the Wilson
coefficients, and $\tau$ represents a combination of CKM matrix
elements: $\tau=-V_{ts}^*V_{td}/V_{ud}V_{us}^*$.  To calculate the
decay amplitudes $A_2$ and $A_0$, we need to calculate the matrix
elements $\langle\pi\pi|Q_i|K^0\rangle$ on the lattice.

\begin{figure*}
\begin{ruledtabular}
 \begin{tabular}{cc}
\includegraphics[width=0.4\textwidth]{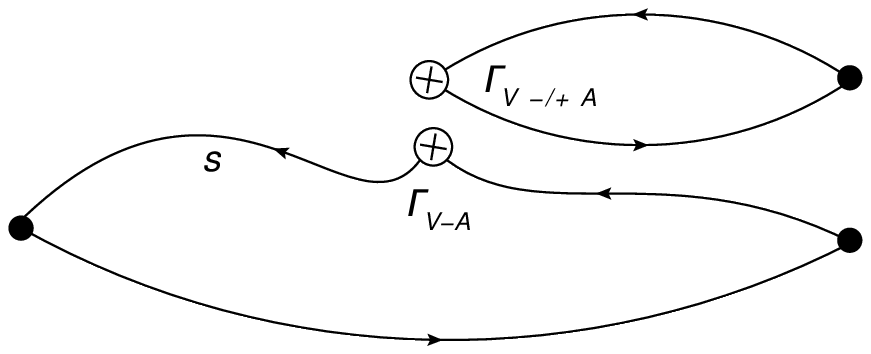}&
\includegraphics[width=0.4\textwidth]{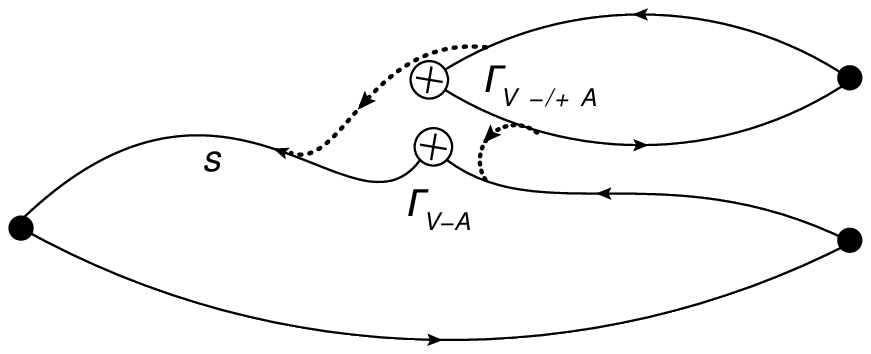}\\
\fig{1}/\fig{3} & \fig{2}/\fig{4}\\
\hline
\includegraphics[width=0.4\textwidth]{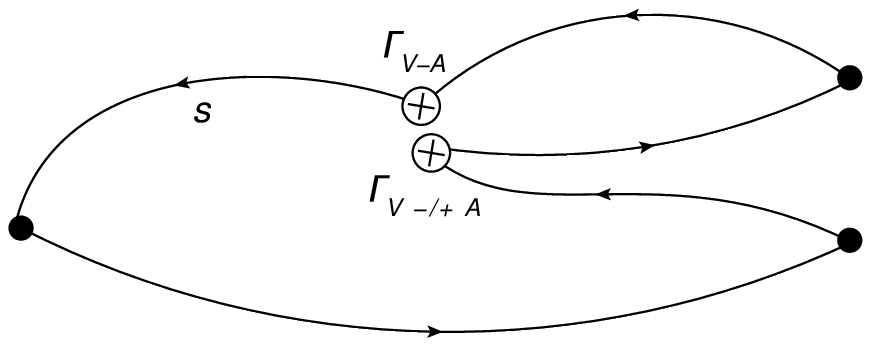}&
\includegraphics[width=0.4\textwidth]{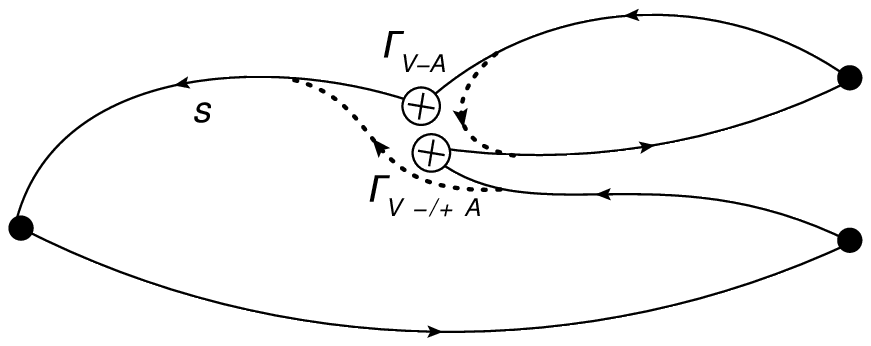}\\
\fig{5}/\fig{7} & \fig{6}/\fig{8}\\
\end{tabular}
\end{ruledtabular}
\caption{Diagrams representing the eight $K^0\rightarrow\pi\pi$ contractions
of $type1$, where $\Gamma_{V\pm A}=\gamma_\mu(1\pm\gamma_5)$.  The black
dot indicates a $\gamma_5$ matrix, which is present in each operator
creating or destroying a pseudoscalar meson.}
\label{fig:type1}
\end{figure*}

\begin{figure*}
\begin{ruledtabular}
\begin{tabular}{cc}
\includegraphics[width=0.4\textwidth]{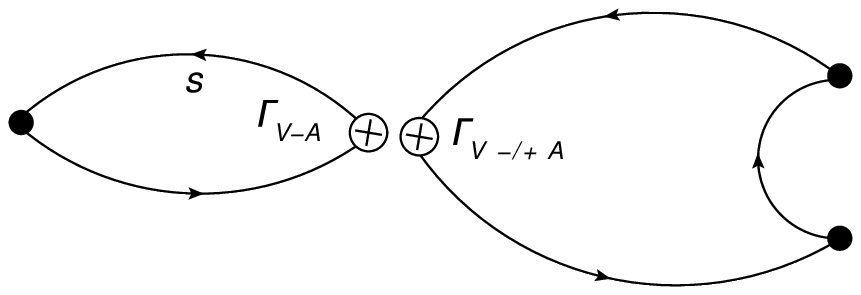}&
\includegraphics[width=0.4\textwidth]{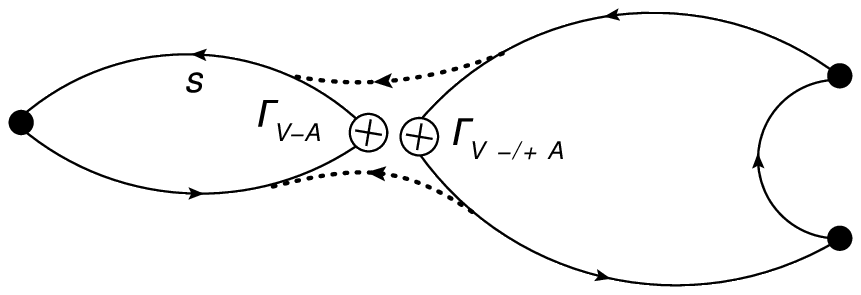}\\
\fig{9}/\fig{11} & \fig{10}/\fig{12}\\
\hline
\includegraphics[width=0.4\textwidth]{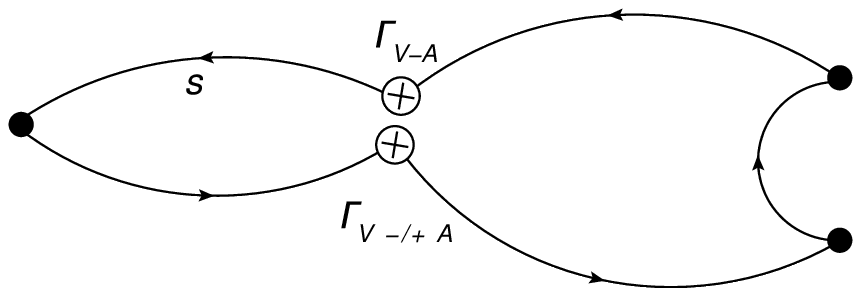}&
\includegraphics[width=0.4\textwidth]{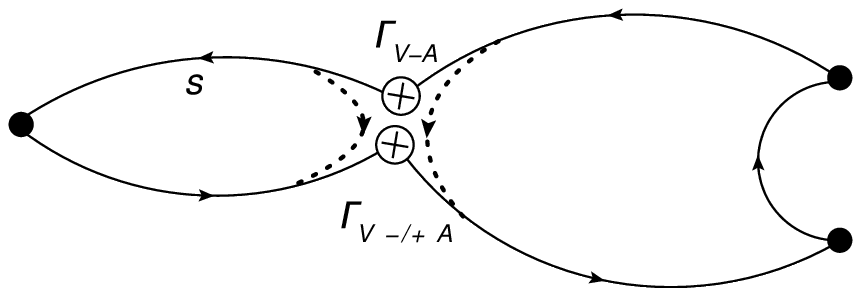}\\
\fig{13}/\fig{15} & \fig{14}/\fig{16}\\
\end{tabular}
\end{ruledtabular}
\caption{Diagrams for the eight $type2$ $K^0\rightarrow\pi\pi$ contractions.}
\label{fig:type2}
\end{figure*}

\begin{figure*}
\begin{ruledtabular}
\begin{tabular}{cc}
\includegraphics[width=0.4\textwidth]{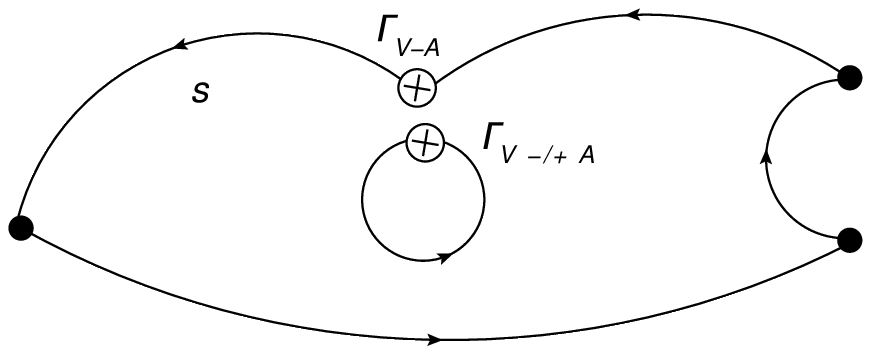}&
\includegraphics[width=0.4\textwidth]{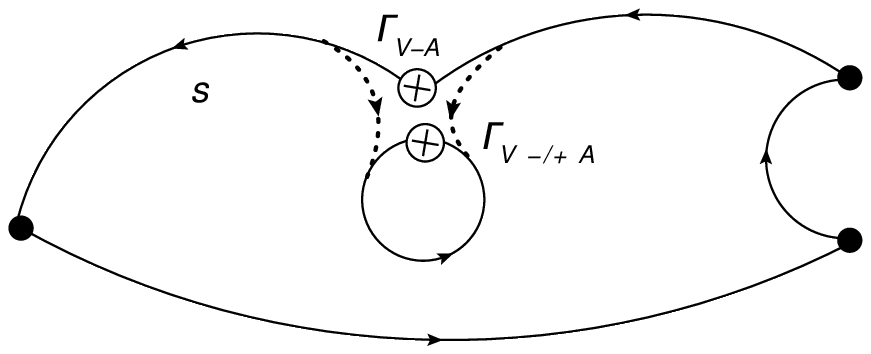}\\
\fig{17}/\fig{19} & \fig{18}/\fig{20} \\
\hline
\includegraphics[width=0.4\textwidth]{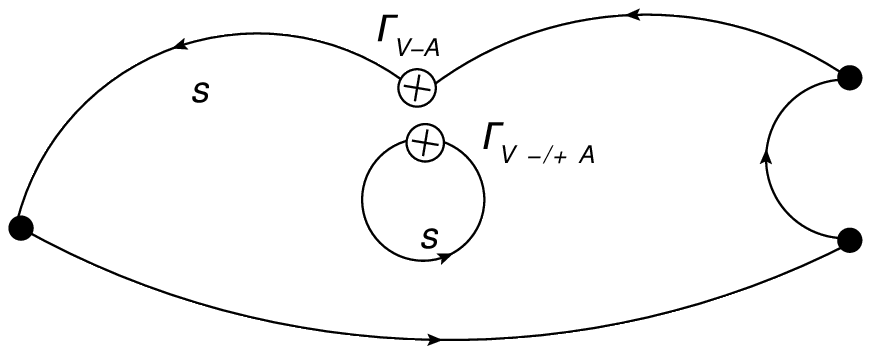}&
\includegraphics[width=0.4\textwidth]{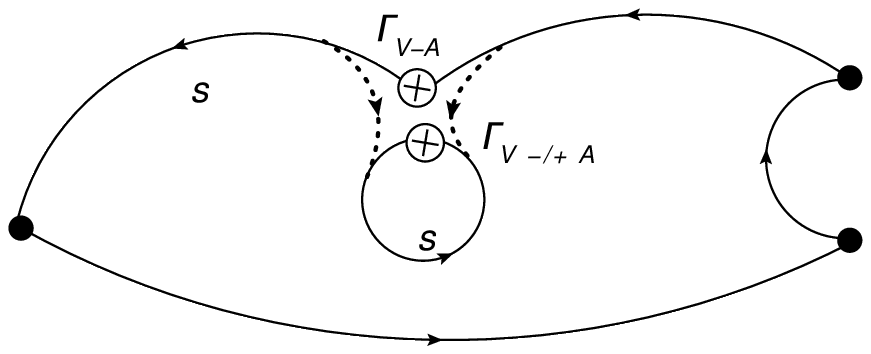}\\
\fig{21}/\fig{23} & \fig{22}/\fig{24} \\
\hline
\includegraphics[width=0.4\textwidth]{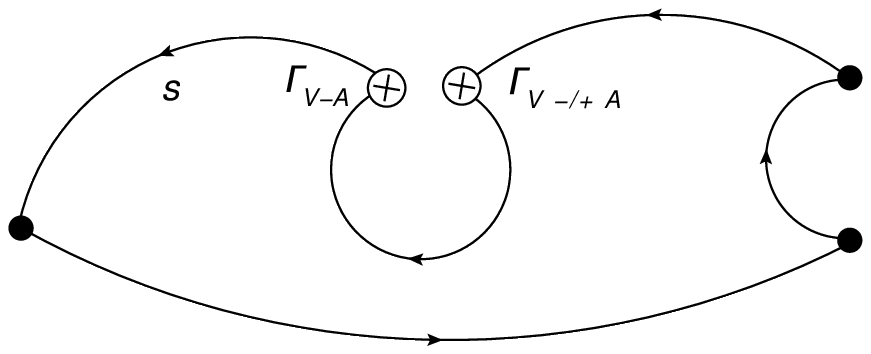}&
\includegraphics[width=0.4\textwidth]{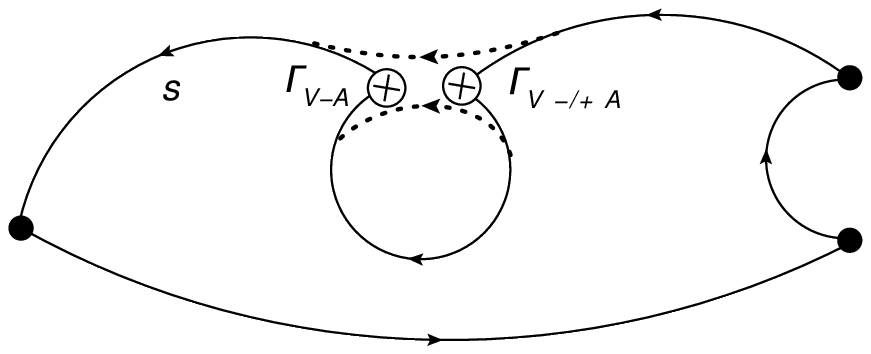}\\
\fig{25}/\fig{27} & \fig{26}/\fig{28} \\
\hline
\includegraphics[width=0.4\textwidth]{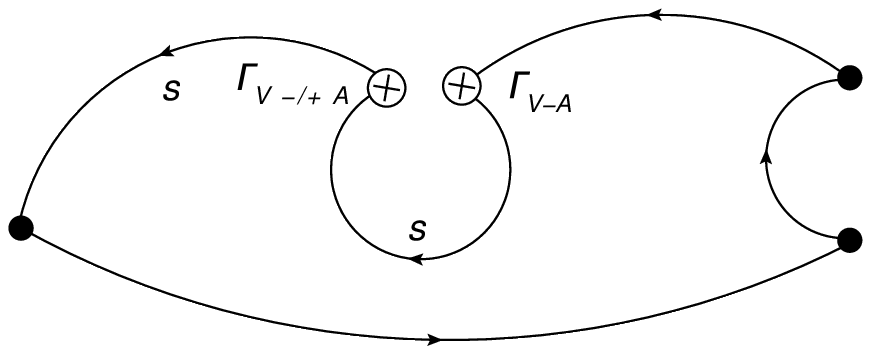}&
\includegraphics[width=0.4\textwidth]{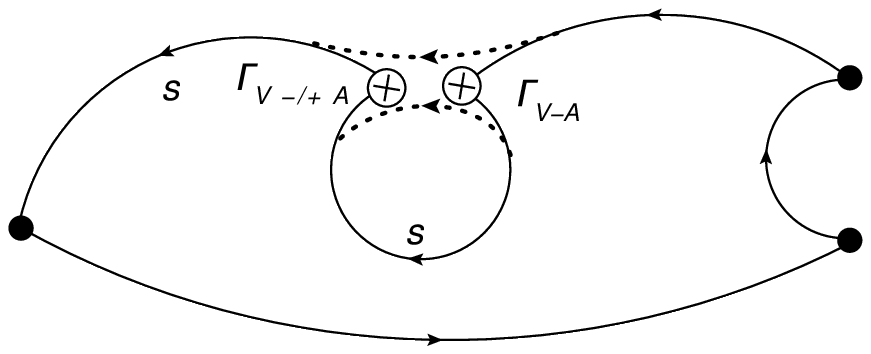}\\
\fig{29}/\fig{31} & \fig{30}/\fig{32} \\
\end{tabular}
\end{ruledtabular}
\caption{Diagrams for the 16  $type3$ $K^0\rightarrow\pi\pi$ contractions.}
\label{fig:type3}
\end{figure*}

\begin{figure*}
\begin{ruledtabular}
\begin{tabular}{cc}
\includegraphics[width=0.4\textwidth]{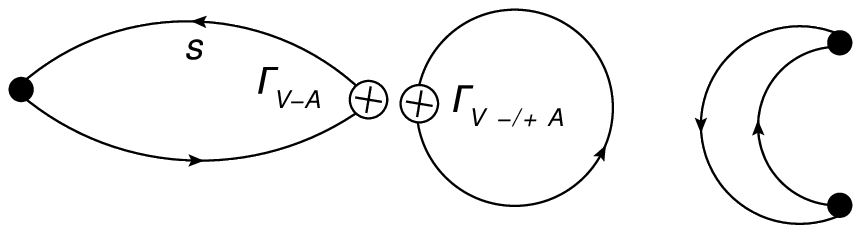}&
\includegraphics[width=0.4\textwidth]{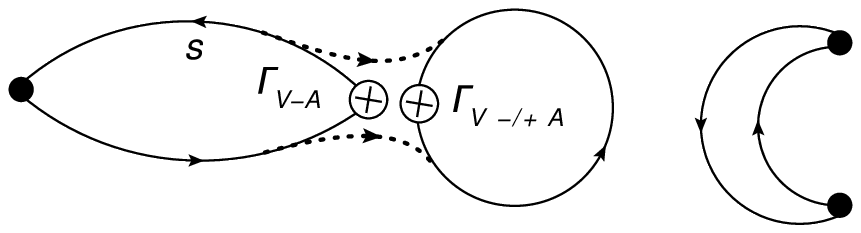}\\
\fig{33}/\fig{35} & \fig{34}/\fig{36} \\
\hline
\includegraphics[width=0.4\textwidth]{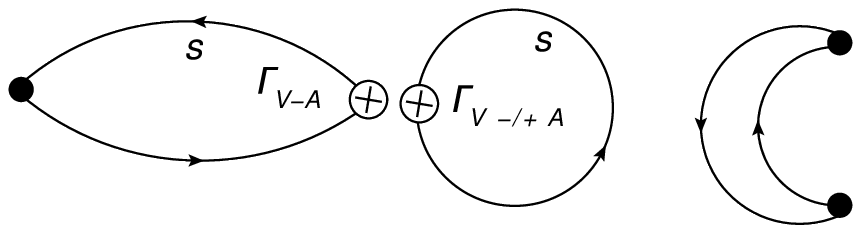}&
\includegraphics[width=0.4\textwidth]{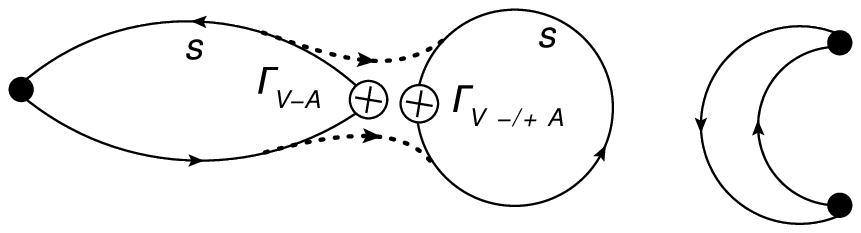}\\
\fig{37}/\fig{39} & \fig{38}/\fig{40} \\
\hline
\includegraphics[width=0.4\textwidth]{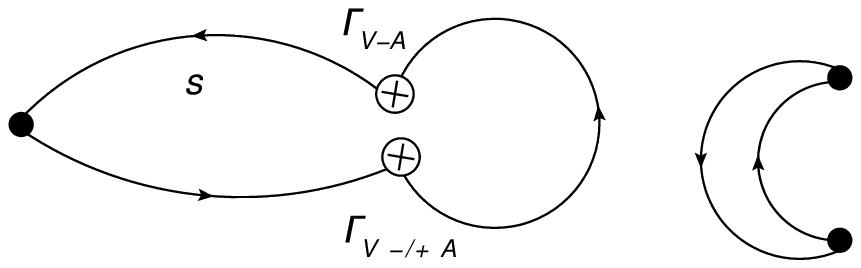}&
\includegraphics[width=0.4\textwidth]{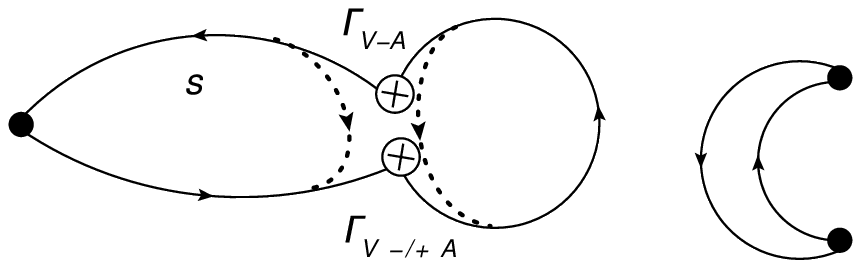}\\
\fig{41}/\fig{43} & \fig{42}/\fig{44} \\
\hline
\includegraphics[width=0.4\textwidth]{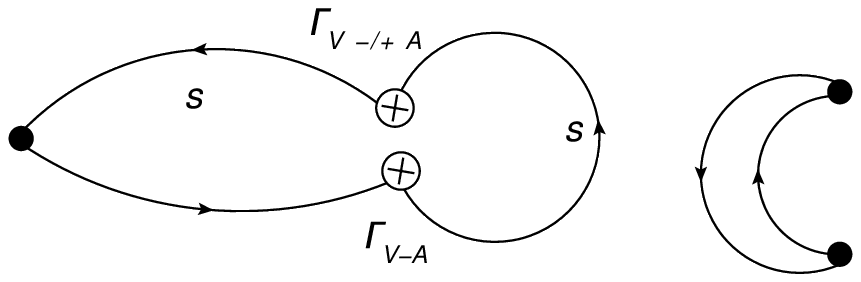}&
\includegraphics[width=0.4\textwidth]{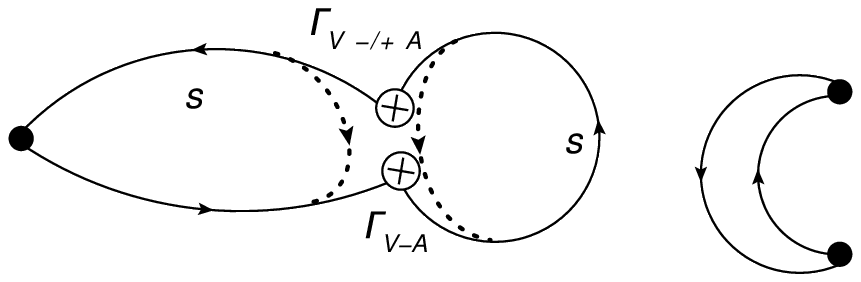}\\
\fig{45}/\fig{47} & \fig{46}/\fig{48} \\
\end{tabular}
\end{ruledtabular}
\caption{Diagrams for the sixteen $type4$ $K^0\rightarrow\pi\pi$  contractions.}
\label{fig:type4}
\end{figure*}

We list all of the possible contractions contributing to the matrix
elements $\left<\pi\pi|Q_i|K^0\right>$ in Figs.~\ref{fig:type1}-\ref{fig:type4}.
There are 48 different contractions which are labeled by circled numbers
ranging from 1 to 48, and grouped into four categories labeled
as $type1$, $type2$, $type3$, and $type4$ according to their topology.
Once we have calculated all of these contractions, the correlation
functions $\left<O^{\pi\pi}_I(t_\pi)Q_i(t_{\rm op})K^0(t_K)\right>$
are then obtained as combinations of these contractions.
In order to simplify the following formulae, we use the amplitude
$A_{I,i}(t_\pi,t,t_K)$ to represent three point function
$\langle O^{\pi\pi}_I(t_\pi)Q_i(t_{\rm op})K(t_K)\rangle$.  Using this
notation, the $I=2$ amplitudes can be written,
\begin{subequations}
\begin{eqnarray}
A_{2,1}(t_\pi,t_{\rm op},t_K) & = &i\sqrt{\frac{2}{3}}\{\fig{1}-\fig{5}\}\\
A_{2,2}(t_\pi,t_{\rm op},t_K) & = &i\sqrt{\frac{2}{3}}\{\fig{2}-\fig{6}\}\\
A_{2,3}(t_\pi,t_{\rm op},t_K) & = &0\\
A_{2,4}(t_\pi,t_{\rm op},t_K) & = &0\\
A_{2,5}(t_\pi,t_{\rm op},t_K) & = &0\\
A_{2,6}(t_\pi,t_{\rm op},t_K) & = &0\\
A_{2,7}(t_\pi,t_{\rm op},t_K) & = &i\sqrt{\frac{3}{2}}\{\fig{3}-\fig{7}\}\\
A_{2,8}(t_\pi,t_{\rm op},t_K) & = &i\sqrt{\frac{3}{2}}\{\fig{4}-\fig{8}\}\\
A_{2,9}(t_\pi,t_{\rm op},t_K) & = &i\sqrt{\frac{3}{2}}\{\fig{1}-\fig{5}\}\\
A_{2,10}(t_\pi,t_{\rm op},t_K) & = &i\sqrt{\frac{3}{2}}\{\fig{2}-\fig{6}\}
\end{eqnarray}
\label{eq:I2}
\end{subequations}

and in the I=0 case,

\begin{widetext}
\begin{subequations}
\begin{eqnarray}
A_{0,1}(t_\pi,t_{\rm op},t_K) & = &i\frac{1}{\sqrt{3}}\{-\fig{1}-2\cdot\fig{5}
     +3\cdot\fig{9}+3\cdot\fig{17}-3\cdot \fig{33}\}\\
A_{0,2}(t_\pi,t_{\rm op},t_K) & = &i\frac{1}{\sqrt{3}}\{-\fig{2}-2\cdot\fig{6}
     +3\cdot\fig{10} +3\cdot\fig{18}-3\cdot\fig{34}\}\\
A_{0,3}(t_\pi,t_{\rm op},t_K) & = &i\sqrt{3}\{-\fig{5}+2\cdot\fig{9}-\fig{13}
     +2\cdot\fig{17}+\fig{21} \\ \nonumber && \hskip 0.5in
     -\fig{25} -\fig{29}-2\cdot\fig{33}-\fig{37} +\fig{41}+\fig{45}\}\\
A_{0,4}(t_\pi,t_{\rm op},t_K) & = &i\sqrt{3}\{-\fig{6}+2\cdot\fig{10}-\fig{14}
     +2\cdot\fig{18}+\fig{22}  \\ \nonumber && \hskip 0.5in
     -\fig{26} -\fig{30}-2\cdot\fig{34}-\fig{38}+\fig{42}+\fig{46}\}\\
A_{0,5}(t_\pi,t_{\rm op},t_K) & = &i\sqrt{3}\{-\fig{7}+2\cdot\fig{11}-\fig{15}
     +2\cdot\fig{19}+\fig{23} \\ \nonumber && \hskip 0.5in
     -\fig{27} -\fig{31}-2\cdot\fig{35}-\fig{39}+\fig{43}+\fig{47}\}\\
A_{0,6}(t_\pi,t_{\rm op},t_K) & = &i\sqrt{3}\{-\fig{8}+2\cdot\fig{12}-\fig{16}
     +2\cdot\fig{20}+\fig{24} \\ \nonumber && \hskip 0.5in
     -\fig{28}-\fig{32}-2\cdot\fig{36}-\fig{40}+\fig{44}+\fig{48}\} \\
A_{0,7}(t_\pi,t_{\rm op},t_K) & = &i\frac{\sqrt{3}}{2}\{-\fig{3}-\fig{7}+\fig{11}
     +\fig{15}+\fig{19} \\ \nonumber && \hskip 0.5in
     -\fig{23}+\fig{27}+\fig{31}-\fig{35}+\fig{39}-\fig{43}-\fig{47}\}\\
A_{0,8}(t_\pi,t_{\rm op},t_K) & = &i\frac{\sqrt{3}}{2}\{-\fig{4}-\fig{8}+\fig{12}
     +\fig{16}+\fig{20} \\ \nonumber && \hskip 0.5in
     -\fig{24}+\fig{28}+\fig{32}-\fig{36}+\fig{40}-\fig{44}-\fig{48}\}\\
A_{0,9}(t_\pi,t_{\rm op},t_K) & = &i\frac{\sqrt{3}}{2}\{-\fig{1}-\fig{5}+\fig{9}
     +\fig{13}+\fig{17} \\ \nonumber && \hskip 0.5in
     -\fig{21}+\fig{25}+\fig{29}-\fig{33}+\fig{37}-\fig{41}-\fig{45}\}\\
A_{0,10}(t_\pi,t_{\rm op},t_K) & = &i\frac{\sqrt{3}}{2}\{-\fig{2}-\fig{6}+\fig{10}
     +\fig{14}+\fig{18} \\ \nonumber && \hskip 0.5in
     -\fig{22}+\fig{26}+\fig{30}-\fig{34}+\fig{38}-\fig{42}-\fig{46}\},
\end{eqnarray}
\label{eq:I0}
\end{subequations}
\end{widetext}
where the factor $i$ comes from our definition of the interpolation
operator for the mesons, {\it e.g.} $K^0 = i(\overline{d}\gamma_5 s)$.

A few notes about the contractions shown in the Figs.~\ref{fig:type1} -
\ref{fig:type4} may be useful:
\begin{enumerate}
\item The contractions identified by circled numbers do not carry the minus
sign required when there is an odd number of fermion loops.  Instead, the
signs are included explicitly in Eqs.~\ref{eq:I2} and \ref{eq:I0}.
\item The routing of the solid line indicates spin contraction while that of
the dashed line indicates the contraction of color indices.  If there is
no dashed line, then solid line indicates connections implied by the trace
over both color and spin indices.  (This will be explained in more detail 
below.)
\item A line represents a light quark propagator if it is not explicitly
labeled with 's'.  Up and down quarks and particular flavors of pion
are not distinguished in Figs.~\ref{fig:type1} - \ref{fig:type4}.  Instead these
specific contractions of strange and light quark propagators are combined
in Eqs.~\ref{eq:I2} and \ref{eq:I0} to give the $I=2$ and $I=0$ amplitudes
directly.
\item Using Fierz symmetry, it can be shown that there are 12 identities
among these contractions:
\begin{subequations}
\begin{eqnarray}
\fig{6}=-\fig{1},  \quad \fig{5}=-\fig{2},   \quad \fig{14}=-\fig{9}, \quad \fig{13}=-\fig{10}, \label{eq:Fierz_a}\\
\fig{26}=-\fig{17},\quad \fig{25}=-\fig{18}, \quad \fig{29}=-\fig{22},\quad \fig{30}=-\fig{21}, \\
\fig{42}=-\fig{33},\quad \fig{41}=-\fig{34}, \quad \fig{45}=-\fig{38},\quad\fig{46}=-\fig{37}.
\end{eqnarray}
\end{subequations}
A consequence of these identities is that Eq.~\ref{eq:I0} is consistent
with only seven of the ten operators $Q_i$ being linearly independent and
with the three usual relations:
\begin{subequations}
\begin{eqnarray}
Q_{10}-Q_9 & = & Q_4-Q_3 \\
Q_4 - Q_3 & = & Q_2 -Q_1 \\
2Q_9 & = & 3Q_1-Q_3.
\end{eqnarray}
\label{eq:identity}
\end{subequations}
\item Based on charge conjugation symmetry and $\gamma^5$ hermiticity,
the gauge field average of each of these contractions is real.
\item The loop contractions of $type3$ and $type4$ are calculated using the
Gaussian, stochastic wall sources described in Sec.~\ref{sec:Comp_details}.
\end{enumerate}

In order to make our approach more explicit, we will discuss some
examples.  First consider the two contractions of $type1$ identified as $\fig{1}$
and $\fig{2}$ and shown in the top half of Fig.~\ref{fig:type1}:
\begin{widetext}
\begin{eqnarray}
\fig{1}&=&\mathrm{Tr}\Bigl\{\gamma_\mu(1-\gamma_5)L(x_{op},t_\pi)
                                           L(x_{op},t_\pi)^\dagger\Bigr\}
\label{eq:type1_example} \\ \nonumber &&\hskip 0.5in \cdot
\mathrm{Tr}\Bigl\{\gamma^\mu(1-\gamma_5)L(x_{op},t_\pi)\gamma^5
          \left[\sum_{\vec x_\pi} L((\vec x_\pi,t_\pi),t_K)\right]
          S(x_{op},t_K)^\dagger\Bigr\} \\
\fig{2}&=&\mathrm{Tr}_c\Biggl\{\mathrm{Tr}_s\Bigl\{\gamma_\mu(1-\gamma_5)L(x_{op},t_\pi)
                                           L(x_{op},t_\pi)^\dagger\Bigr\}
\label{eq:type2_example} \\ \nonumber &&\hskip 0.5in \cdot
\mathrm{Tr}_s\Bigl\{\gamma^\mu(1-\gamma_5)L(x_{op},t_\pi)\gamma_5
          \left[\sum_{\vec x_\pi} L((\vec x_\pi,t_\pi),t_K)\right]
          S(x_{op},t_K)^\dagger\Bigr\}\Biggr\},
\end{eqnarray}
\end{widetext}
where $t_K$ is the time of the kaon wall source, $t_\pi$ the time
at which the two pions are absorbed and $x_{op}=(\vec x_{op},
t_{op})$ the location of the weak operator.  The function
$L(x_{\rm sink}, t_{\rm src})$ is the light quark propagator, a $12 \times 12$
spin-color matrix, while $S(x_{\rm sink},t_{\rm src})$ is the strange quark
propagator.  The hermitian conjugation operation, $\dagger$, operates on
these $12 \times 12$ matrices.   We use Tr$_c$ to indicate a color trace,
Tr$_s$ a spin trace, and Tr, with no subscript, stands for both a spin
and color trace.  We have also used the $\gamma^5$ hermiticity of the
quark propagators to realize the combination of quark propagators given
in Eqs.~\ref{eq:type1_example} and \ref{eq:type2_example}, allowing
both contractions to be constructed from light and strange propagators
computed using Coulomb gauge fixed wall sources located only at the times
$t_\pi$ and $t_K$.  Note the sum over the spatial components of the sink
$\vec x_\pi$ creates a symmetrical wall sink provided that the appropriate
Coulomb gauge transformation matrix has been applied to the sink color
index of this propagator to duplicate the Coulomb gauge transformation
that was used to create the Coulomb gauge fixed wall source.
We will sum over the spatial location, $\vec x_{op}$, of the weak
operator, to project onto zero spatial momentum and improve statistics.
Below we will show results as a function of the separations between
$t_\pi$, $t_{\rm op}$ and $t_K$.

As a third example, which illustrates the use of random wall sources,
consider contraction $\fig{19}$ shown in Fig.~\ref{fig:type3}.  Using the
notation introduced above, this contraction is given by
\begin{widetext}
\begin{eqnarray}
\fig{19}&=&\mathrm{Tr}\Bigl\{\gamma_\mu(1+\gamma_5)L^R(x_{\rm op},t_{\rm op})
                           \Bigr\}\eta(x_{\rm op})^*
\label{eq:type19_example} \\ \nonumber &&\hskip 0.1in \cdot
\mathrm{Tr}\Bigl\{\gamma^\mu(1-\gamma_5)L(x_{\rm op},t_\pi)
      \Biggl[\sum_{\vec x_\pi'} L\Bigl((\vec x_\pi',t_\pi),t_\pi\Bigr)^\dagger\Biggr]
      \Biggl[\sum_{\vec x_\pi}  L\Bigl((\vec x_\pi, t_\pi),t_K\Bigr)\Biggr]
      S(x_{\rm op},t_K)^\dagger\Bigr\}.
\end{eqnarray}
\end{widetext}
Here $\eta(x)$ is the value of the complex, Gaussian random wall source at
the space-time position $x$, while $L^R(x_{\rm sink},t_{\rm src})$ is the propagator
whose source is $\eta(x) \delta(x_0-t_{\rm src})$.  The Dirac delta function
$\delta(x_0-t_{\rm src})$ restricts the source to the time plane $t=t_{\rm src}$.
In the usual way, the average over the random source $\eta(\vec x)$ which
accompanies the configuration average, will set to zero all terms in which
the source and sink positions for the propagator $L^R(x_{\rm op},t_{\rm op})$ in
Eq.~\ref{eq:type19_example} differ, giving us the contraction implied by the
closed loop in the top left panel of Fig.~\ref{fig:type3}.  By using 32
separate propagators each with a random source non-zero on only one of our
32 time slices we obtain more statistically accurate results than would
result from a single random source spread over all times.

An important objective of this calculation is to learn how to accurately evaluate
the quark loop integration that is present in $type3$ and $type4$ graphs
and which contains a $1/a^2$, quadratically divergent component.  As can be
recognized from the structure of the diagrams, these divergent terms can be
interpreted as arising from the mixing between the dimension-six operators
$Q_i$ (for all $i$ but 7 and 8) and a dimension-3 ``mass'' operator of the
form $\overline{s}\gamma_5d$.  Such divergent terms are expected and do not
represent a breakdown of the standard effective Hamiltonian written in
Eq.~\ref{eq:weak_eff}.  In fact, given the good chiral symmetry of domain
wall fermions all other operators with dimension less than six which might
potentially mix with those in Eq.~\ref{eq:weak_eff} will vanish if the
equations of motion are imposed.  Therefore these operators cannot contribute
to the Green's functions evaluated in Eqs.~\ref{eq:I2} and \ref{eq:I0} where
the operators in $H_W$ are separated in space-time from those operators
creating the $K$ meson and destroying the $\pi$ mesons, a circumstance in
which the equations of motion can be applied.

The problematic operator $\overline{s}\gamma_5d$ is not explictly removed
from the effective Hamiltonian because, again using the equations of
motion, $\overline{s}\gamma_5d$ can be written as the divergence of an axial
current and hence will vanish in the physical case where the weak operator
$H_W$ carries no four-momentum and is evaluated between on-shell states.
While we can explicitly sum the effective Hamiltonian density ${\cal H}_W$
over space to ensure $H_W$ carries no spatial momentum, to ensure that no energy
is transferred we must arrange that the kaon mass and two-pion energy are
equal.  We may achieve this condition, at least approximately, but there
will be contributions from heavier states, which are normally exponentially suppressed,
but which will violate energy conservation and hence will be enhanced by this divergent
$\overline{s}\gamma_5d$ term.

Since $\overline{s}\gamma_5d$ will not contribute to the physical,
energy-conserving $K \rightarrow \pi\pi$ amplitude, there is no theoretical
requirement that it be removed.  The coefficient of this
$\overline{s}\gamma_5d$ piece is both regulator dependent and irrelevant.
The contribution of these terms in a lattice calculation of
$K \rightarrow \pi\pi$ decay amplitudes will ultimately vanish as the equality of the initial and final energies is made more precise and as increased time separations are achieved. 
However, the unphysical effects of this $\overline{s}\gamma_5d$
mixing are much more easily suppressed by reducing the size of this
irrelevant term than by dramatically increasing the lattice size and
collecting the substantially increased statistics required to work at
large time separations.

A direct way to remove this $1/a^2$ enhancement is to explicitly subtract
an $\alpha_i\overline{s}\gamma_5d$ term from each of the relevant operators
$Q_i$ where the coefficient $\alpha_i$ can be fixed by imposing the condition:
\begin{equation}
\left<0| Q_i - \alpha_i \overline{s}\gamma_5d|K\right>=0,
\label{eq:subtraction_condition}
\end{equation}
a condition that is typically required in the chiral perturbation theory for
$K \rightarrow \pi\pi$~\cite{Blum:2001xb}.  
Of course, this arbitrary condition
will leave a finite, regulator-dependent $\overline{s}\gamma_5d$ piece
behind in the subtracted operator $Q_i  - \alpha_i \overline{s}\gamma_5d$.
However, this unphysical piece will not contribute to the energy-conserving
amplitude being evaluated.  Since it is no longer $1/a^2$-enhanced its effects
on our calculation will be similar to those of the many other energy non-conserving
terms which we must suppress by choosing equal energy $K$ and $\pi \pi$ states
and using sufficient large time separation to suppress the contributions
of excited states.

Following Eq.~\ref{eq:subtraction_condition} we will choose the coefficient
$\alpha_i$ from the ratio
\begin{equation}
\alpha_i = \frac{\left<0|Q_i|K^0\right>}{\left<0|\overline{s}\gamma_5d|K^0\right>}.
\end{equation}
(Note, with this definition the coefficient $\alpha_i$ is proportional to the
difference of the strange and light quark masses.)  Thus, we will improve
the accuracy when calculating graphs of $type3$ and $type4$ by including an
explicit subtraction term for those operators $Q_i$ where mixing with
$\overline{s}\gamma_5d$ is permitted by the symmetries (all but $Q_7$
and $Q_8$):
\begin{equation}
\left<O^{\pi\pi}_0(t_\pi)Q_i(t_{\rm op})K^0(t_K)\right>_{sub}
  = \left<O^{\pi\pi}_0(t_\pi)Q_i(t_{\rm op})K^0(t_K)\right>
  - \alpha_i \left<O^{\pi\pi}_0(t_\pi)\overline{s}\gamma_5d(t_{\rm op})
           K^0(t_K)\right>.
\label{eq:sub}
\end{equation}
We should recognize that there is a second, divergent, parity-even operator
$\overline{s}d$ which mixes with our operators $Q_i$.  However, we choose to
neglect this effect because parity symmetry prevents it from contributing to
either the $K \rightarrow \pi\pi$ or $K \rightarrow |0\rangle$ correlation
functions being evaluated here.

\begin{figure*}
\begin{ruledtabular}
\begin{tabular}{cc}
\includegraphics[width=0.4\textwidth]{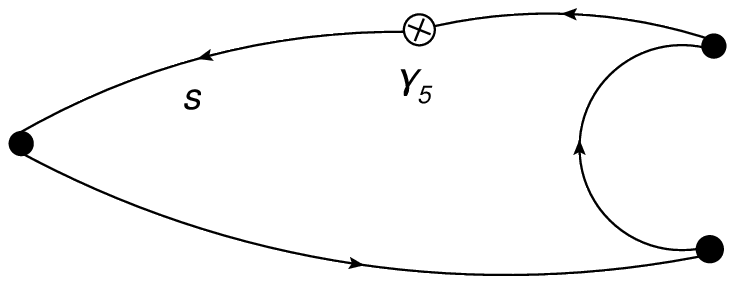} &
\includegraphics[width=0.4\textwidth]{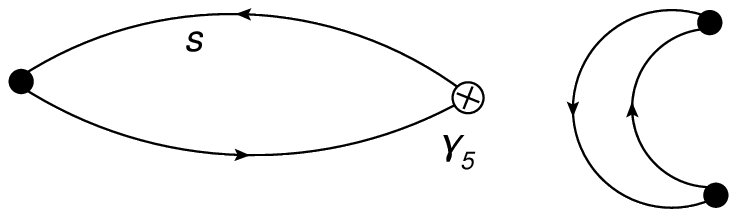} \\
$mix3$ & $mix4$
\end{tabular}
\end{ruledtabular}
\caption{Diagrams showing the contractions needed to evaluate the subtraction
terms.  These are labeled $mix3$ and $mix4$ and constructed from the $type3$
and $type4$ contractions by replacing the operator $Q_i$ and fermion
loop with the vertex $\overline{s}\gamma_5d$.}
\label{fig:s5d}
\end{figure*}

The amplitude $\left<O^{\pi\pi}_0(t_\pi)\overline{s}\gamma_5d(t_{\rm op})
K^0(t_K)\right>$ includes two contractions, one connected and one
disconnected as shown in Fig.~\ref{fig:s5d}.  These terms, which arise
from the mixing of the operators $Q_i$ with $\overline{s}\gamma_5d$, are
labeled $mix3$ and $mix4$.  To better visualize the contributions from
different types of contractions, we can write the right hand side of
Eq.~\ref{eq:sub} symbolically as
\begin{eqnarray}
& & type1 + type2 +type3 + type4 - \alpha\cdot(mix3 + mix4) \nonumber \\
& = & type1 + type2 + sub3 + sub4,
\end{eqnarray}
where $sub3=type3-\alpha\cdot mix3$ and $sub4=type4-\alpha\cdot mix4$.
Note, here and in later discussions we refer to the term being subtracted as
``mix'' and the final difference as the subtracted amplitude ``sub''.

\section{$K^0 \rightarrow \pi\pi$ $\Delta I=3/2$ amplitude}
\label{sec:DeltaI_3-2}

\begin{figure*}
\begin{tabular}{cc}
\includegraphics[width=0.5\textwidth]{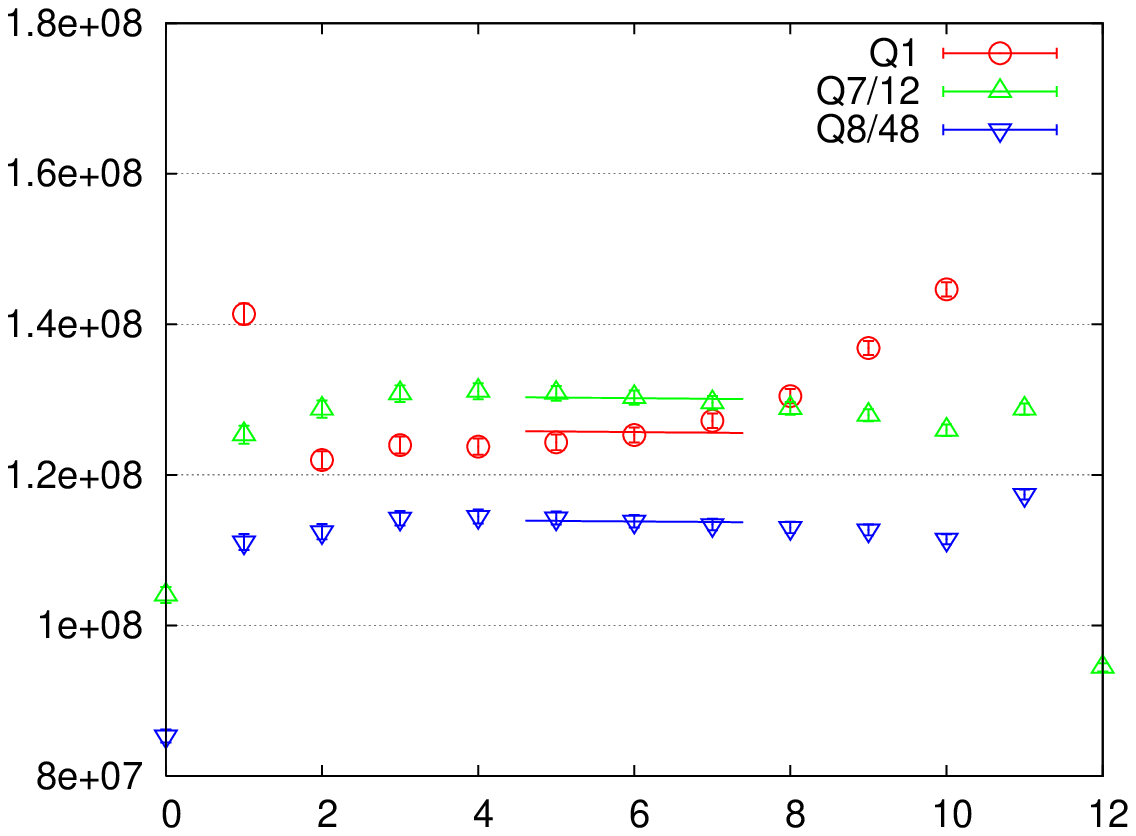} &
\includegraphics[width=0.5\textwidth]{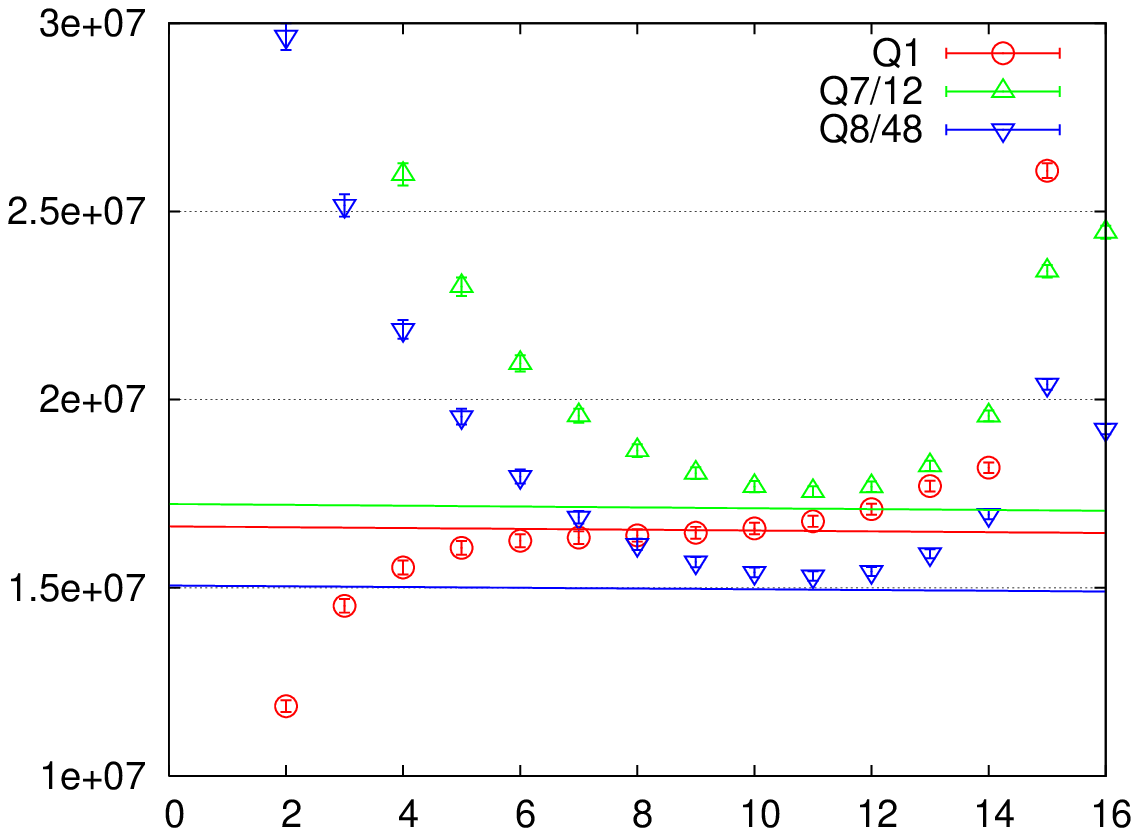} \\
$\Delta=12$ & $\Delta=16$
\end{tabular}
\caption{Plots of the $\Delta I=3/2$ $K^0\rightarrow\pi-\pi$ correlation
functions for kaon source and $\pi-\pi$ sink separations of $\Delta=12$ (left
panel) and $16$ (right panel).  The $x$-axis gives the time $t$ specifying
the time slice over which the operator, $Q_i(\vec x,t)$, $i=1,$ 7, 8, is averaged.
The results for the operator $Q_7$ are divided by 12, and those for $Q_8$ by
48 to allow the results to be shown in the same graph.   The correlators
$C_{2,i}(\Delta,t)$ are fit using the $\Delta=12$ data with a fitting range
$5 \le t \le 7$.  The resulting constants are shown as horizontal lines in
both the $\Delta=12$ and 16 graphs.  We can see that the $\Delta=16$ data are
consistent with those from $\Delta=12$, but receive large contributions
from the around-the-world paths.}
\label{fig:I2}
\end{figure*}

As Eqs.~\ref{eq:I2} and \ref{eq:Fierz_a} show, the $\Delta I=3/2$
$K^0 \rightarrow 2\pi$ decay amplitude includes only $type1$ contractions
and four of the correlation functions are related
\begin{equation}
A_{2,10} = A_{2,9} = \frac{3}{2}A_{2,1} = \frac{3}{2}A_{2,2}.
\end{equation}
Therefore, we need only to calculate $A_{2,1}$, $A_{2,7}$ and $A_{2,8}$.
The corresponding three correlation functions, $C_{2,i}(\Delta, t)$ for $i=1$,
7 and 8, with the choice of $m_K^{(1)}$ for the kaon mass, are shown in Fig.~\ref{fig:I2}.  Here we exploit our propagator calculation for sources
on each of the 32 time slices to compute $C_{2,i}(\Delta, t)$ from an average
over all 32 source positions:
\begin{equation}
C_{2,i}(\Delta, t) =
  \frac{1}{32}\sum_{t'=0}^{31} A_{2,i}(t_\pi=t'+\Delta,t_{\rm op}=t+t',t_K=t').
\label{eq:I_2_correlator}
\end{equation}
In Fig.~\ref{fig:I2} we plot  $C_{2,i}(\Delta, t)$ for $0 < t < \Delta$
at fixed $\Delta = 12$ or 16.  Table~\ref{tab:mass} shows that $m_K^{(1)}$
is almost equal to the energy of $I=2$, $\pi-\pi$ state, so the 3-point
correlation function $C_{2,i}(\Delta, t)$ should be approximately
independent of $t$ in the central region where the time coordinate
of the operator is far from both the kaon and the two-pion sources,
$0 \ll t \ll \Delta$.

We fit the correlators $C_{2,i}(\Delta, t)$ using a single free parameter
$M_i^{3/2,{\rm lat}}$:
\begin{eqnarray}
C_{2,i}(\Delta, t)& = & M_i^{3/2,{\rm lat}} N_{\pi\pi} N_K
                   e^{-E_{\pi\pi}\Delta} e^{-(m_K-E_{\pi\pi})t},
\label{eq:I_2_fit}
\end{eqnarray}
where $N_K$, $m_K$ and $N_{\pi\pi}$, $E_{\pi\pi}$ are determined by
fitting the kaon and two-pion correlators respectively:
\begin{eqnarray}
\frac{1}{32}\sum_{t'=0}^{31}\left< K(t+t') K(t') \right>
   &=& N_K^2\left( e^{-m_Kt} + e^{-m_K(T-t)}\right) \\
\frac{1}{32}\sum_{t'=0}^{31}\left< O^{\pi\pi}_2(t+t') O^{\pi\pi}_2(t') \right>
   &=& N_{\pi\pi}^2\left( e^{-E_{\pi\pi} t} + e^{-E_{\pi\pi}(T-t)} + C \right).
\end{eqnarray}
The constant $C$ arises when the two pions join the source at $t'$ and
sink at $t+t'$ by traveling in opposite time directions as discussed below.
The fitted results for the matrix elements $M_i^{3/2,{\rm lat}}$ from $\Delta=12$
are listed in Tab.~\ref{tab:M_3/2_lat} in lattice units.

\begin{table}
\caption{Results for the lattice $\Delta I = 3/2$, $K\rightarrow\pi\pi$
transition amplitudes obtained from fitting the 3-point correlation
functions to the functional form given in Eq.~\protect{\ref{eq:I_2_fit}}
for the six operators with $\Delta I=3/2$ components.  The second
column gives the lattice matrix elements $M_i^{3/2,{\rm lat}}(\times 10^{-2})$
while the third and fourth column give their contributions to the real
and imaginary parts of $A_2$.}
\label{tab:M_3/2_lat}
\begin{ruledtabular}
\begin{tabular}{llll}
i     & $M_i^{3/2,{\rm lat}}(\times 10^{-2})$
                   &  Re$(A_2)$(GeV) 
                                    & Im$(A_2)$(GeV)\\
\hline
1     & 0.4892(16) & -1.737(11)e-08 & 0             \\
2     & $=M_1$     & 6.665(42)e-08  & 0             \\
7     & 6.080(18)  & 2.422(16)e-11  & 4.070(26)e-14 \\
8     & 21.26(6)   & -1.979(13)e-10 & -9.646(61)e-12 \\
9     & =1.5$M_1$  & -7.917(50)e-15  & 5.185(24)e-13 \\
10    & =1.5$M_1$  & 6.103(38)e-12  & -1.448(9)e-13 \\
\hline                     	                    
Total & -          & 4.911(31)e-08  & -5.502(40)e-13 \\
\end{tabular}
\end{ruledtabular}
\end{table}

\begin{figure}
\begin{tabular}{c}
\includegraphics[width=0.4\textwidth]{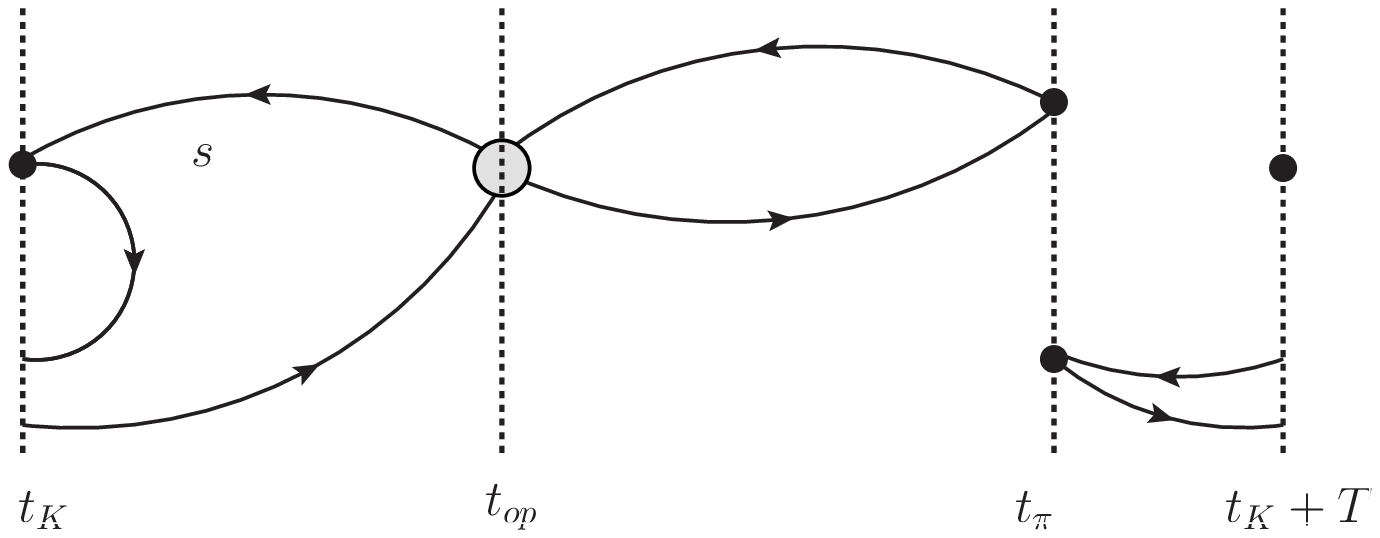} \\
\includegraphics[width=0.4\textwidth]{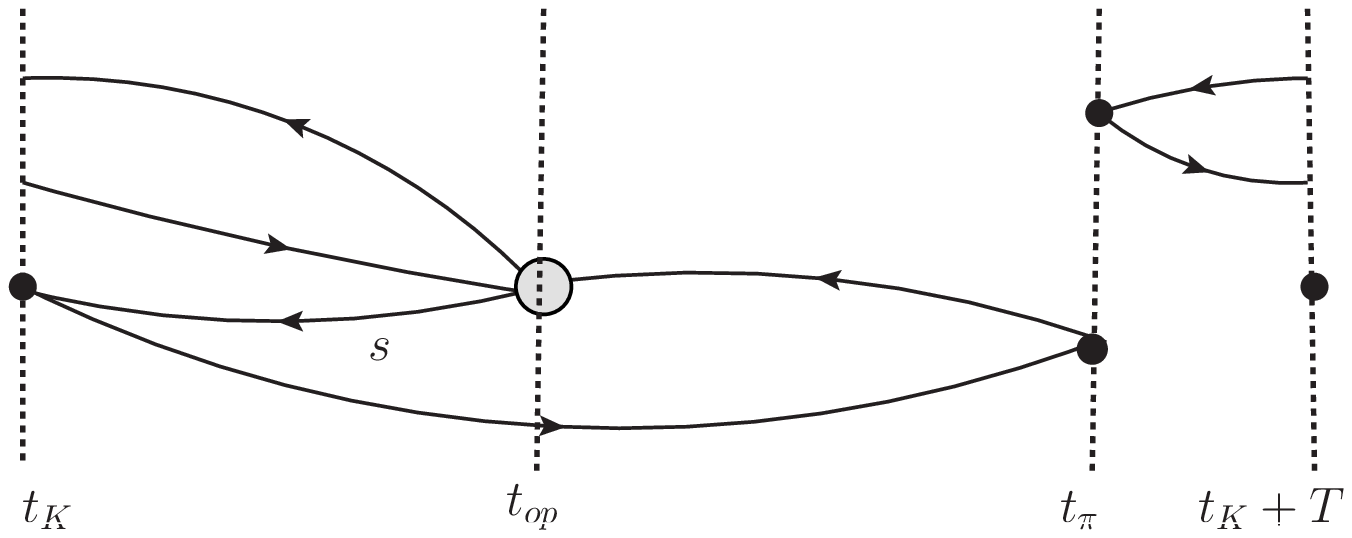}
\end{tabular}
\caption{Diagrams showing the dominant around-the-world paths
contributing to graphs of $type1$. The space-time region between
the kaon wall source at $t_K$ and its periodic recurrence at $t_K+T$
is shown, where $T=32$ is the extent of the periodic lattice
in the time direction.  For this around-the-world path, one pion
travels directly from the pion wall source at $t_\pi$ to the weak
operator, represented by the grey dot at $t_{\rm op}$.  However,
the second pion propagates in the other direction in time, passes
through the periodic boundary and combines with the kaon before
reaching the weak operator at $t_{\rm op}$.}
\label{fig:type1round}
\end{figure}

Figure~\ref{fig:I2} shows that for the operators $Q_7$ and $Q_8$ the
larger separation, $\Delta=16$, between the kaon
source and $\pi-\pi$ sink gives a much shorter plateau region than
the case $\Delta =12$.  This behavior is inconsistent with the usual
expectation that it is the contributions from excited states of the kaon
and pion, contributions which should be suppressed for larger $\Delta$,
that cause the poor plateau.  An alternative, consistent explanation
attributes the shortened plateau region seen for $\Delta =16$ to the
`around-the-world' effect.  This is the contribution to the correlation
function in which the two-pion interpolating operator at the sink
annihilates one pion and creates another (instead of annihilating two
pions as in the $K\to\pi\pi$ contribution we are seeking) and the
process at the weak operator is $K\pi\to\pi$ (instead of $K\to\pi\pi$).
While one pion travels from the weak operator to the $\pi-\pi$ sink
the second is created at the sink and travels forward in time, passing
through the periodic boundary to reach the weak operator together with
the kaon.  The corresponding dominant path is shown in
Fig.~\ref{fig:type1round}.  The time dependence of this behavior can
be estimated as
\begin{equation}
\sim M_i^{3/2,{\rm lat}}N_\pi^2 N_K e^{-m_\pi T} e^{-(E_{K\pi}-m_\pi)t}
\label{eq:round}
\end{equation}
which is $\Delta$ independent but suppressed by the factor
$\exp(-m_\pi T)$, where $N_\pi$ is the analogue of $N_K$ for the case
of single pion production and $T=32$ is the temporal extent of the
lattice.  In contrast, the physical contribution in Eq.~\ref{eq:I_2_fit}
is suppressed by $\exp(-E_{\pi\pi}\Delta)$. Thus, the second, standard
term falls with increasing $\Delta$ and the two
factors are of similar size when $\Delta=T/2$.  Therefore, we should
expect to see a large contamination from such around-the-world effects
in the $\Delta=16$ case, consistent with Fig.~\ref{fig:I2}. In both
panels of that figure, we plot as three horizontal lines the fitted
result from $\Delta=12$ for the three amplitudes $M^{3/2,{\rm lat}}_i
N_{\pi\pi}N_K \exp{-\Delta E_{\pi\pi}}$ for $i=1$, 7 and 8.  The
agreement between these lines and the short plateaus seen in the
right-hand, $\Delta=16$ panel indicates consistency between these
two values of $\Delta$.

Additional evidence supporting this explanation for the short plateau in
the case of $\Delta = 16$ can be obtained by examining the explicit
dependence on $t$ given by Eq.~\ref{eq:round} for the around-the-world
contribution.  Examining the exponential decay with $t$ in the
$\Delta=16$ correlators plotted in the right panel of Fig.~\ref{fig:I2},
for operators $Q_7$ and $Q_8$ we find a value for $E_{K\pi}-m_\pi$
varying between 0.4 and 0.5 depending on the choice of fit range.  A
more accurate value of 0.498(2) can be obtained by fitting the
corresponding correlator for $\Delta =20$ and a fit range of 5 to 11.
The strangeness-carrying state whose mass we have labeled $E_{K\pi}$ can be formed
from two quarks and must be parity even.  Direct calculation of
$E_{K\pi}$ from a scalar $\overline{s} d$ correlator yields
$E_{K\pi}=0.752(12)$ which is consistent with the sum of the result
above, $E_{K\pi}-m_\pi = 0.498(2)$, and the pion mass $m_\pi=0.2437(5)$.
(This energy difference is also close to the kaon mass $m_K^{(1)}=0.50729$
given in Tab.~\ref{tab:mass}.)  Thus, the time dependence expected
from the around-the-world path is quite consistent with that seen
in Fig.~\ref{fig:I2}.

We conclude that it is important to increase the lattice extent in
the time direction both to suppress this around-the-world effect
and to permit the use of a larger source-sink separation giving a
longer plateau.  We will return to discussion of the around-the-world
effect below for the $\Delta I=1/2$ kaon decay where it creates
even greater difficulties.   However, here we can begin to
appreciate the severity of this effect in the $K^0\rightarrow\pi\pi$
system for our temporal lattice extent of 32, given our values of the
lattice spacing and meson masses.

The Wilson coefficients and operators which appear in Eq.~\ref{eq:weak_eff}
are typically expressed in the $\overline{\mbox{MS}}$ scheme.  Thus,
we must change the normalization of our lattice operators $Q_i$ to
that of the $\overline{\mbox{MS}}$ scheme.   We begin by converting
our bare lattice operators into the regularization invariant momentum
(RI/MOM) scheme of Ref.~\cite{Martinelli:1995ty}.  Here we use the earlier 
results of Ref.~\cite{Li:2008zz} which were obtained for the present 
lattice action using the methods of Ref.~\cite{Blum:2001xb}.   In this 
previous work off-shell, Landau-gauge-fixed Green's functions containing 
the lattice operators $Q_i$ are evaluated at specific external momenta
characterized by an energy scale $\mu$.  These results determine a
renormalization matrix $Z_{ij}^\mathrm{RI}(\mu,a)$ which can be used to 
convert the lattice normalization into that of the RI scheme:
\begin{equation}
Q^\mathrm{RI}(\mu)_i = \sum_{j=1}^7 Z_{ij}^{{\rm lat}\to{\rm RI}}(\mu,a) Q_j'.
\label{eq:lat2RI}
\end{equation}
As explained in Appendix A, these equalities hold only when the 
operators appear in physical matrix elements.  The indices $i$ and 
$j$ take on seven values corresponding to the seven independent 
operators in what will be called the chiral basis.  (The primes in this 
equation indicate lattice operators defined in that basis.)   This is 
referred to as nonperturbative renormalization (NPR) because the matrix 
$Z_{ij}^{{\rm lat}\to{\rm RI}}(\mu,a)$ is computed using a lattice 
evaluation of off-shell Green's functions and perturbation theory 
is not used.

Next these $Q^\mathrm{RI}(\mu)_i$ operators are converted to the 
$\overline{\mbox{MS}}$ scheme in which the Wilson coefficients are 
evaluated by applying a conversion matrix 
$R^{\mathrm{RI}\rightarrow\overline{\mathrm{MS}}}_{ij}$ discussed in
detail in Ref.~\cite{Lehner:2011fz}.  Finally the matrix elements of these 
$\overline{\mbox{MS}}$ operators are combined with the Wilson coefficients 
obtained in the $\overline{\mbox{MS}}$ scheme~\cite{Buchalla:1995vs}
using the scale $\mu = 2.15$ GeV to determine the results given later in 
this section for the $\Delta I = 3/2$ amplitude $A_2$ and in the following 
section for the $\Delta I = 1/2$ $A_0$.  These procedures are described in 
greater detail in Appendix A. 

A good approximation to the infinite volume decay amplitude can be
obtained by including the Lellouch-L\"uscher factor ($F$)~\cite{Lellouch:2000pv} 
which relates the $K\rightarrow\pi\pi$ matrix element $M$ of the effective 
weak Hamiltonian of Eq.~\ref{eq:weak_eff} calculated using finite volume 
states normalized to unity to the infinite volume amplitude $A$: 
$|A|^2 = F^2 M^2$ where
\begin{equation}
F^2 = 4\pi\left(\frac{E_{\pi\pi}^2 m_K}{p^3}\right)
          \left\{p\frac{\partial\delta_2(p)}{\partial p}
             + q\frac{\partial\phi(q)}{\partial q}\right\}.
\label{eq:LL}
\end{equation}
Here $p$ is defined through $E_{\pi\pi}=2\sqrt{m_\pi^2+p^2}$, $q=Lp/2\pi$
and $\delta_2(p)$ is the $s$-wave, $I=2$, $\pi-\pi$ scattering phase shift
for pion relative momentum $p$.  The function $\phi(q)$ is known analytically
and given, for example, in Ref.~\cite{Lellouch:2000pv}.  The $I=2$ phase
shift $\delta_2(p)$ is determined from the measured two-pion energy
$E_{\pi\pi}=0.443(13)$ given in Tab.~\ref{tab:mass} and the finite volume
quantization condition~\cite{Luscher:1990ux}
\begin{equation}
\phi(q)+\delta_2(p) = n\pi.
\label{eq:FV_quant}
\end{equation}
For our threshold case we set the integer $n$ to zero and obtain
$\delta_2(p)=-0.0849(43)$.  Because of the small value of $p$ we assume
that $\delta_2(p)$ is a linear homogenous function of $p$ and write
$\delta_2(p)=p \partial\delta_2(p)/\partial p$, the quantity required in
Eq.~\ref{eq:LL} and given in Tab.~\ref{tab:LL}.  (Equation~\ref{eq:LL}
differs by a factor of two from the expression given in the Lellouch-L\"uscher
paper because of our different conventions for the decay amplitude $A$.  With
our conventions the experimental value of Re$(A_2) = 1.48\times 10^{-8}$ GeV.)

In the limit of non-interacting pions, the factor $F$ becomes $F_\mathrm{free}^2 =
2(2m_\pi)^2 m_K L^3$, which reflects the different normalization of states
in a box and plane wave states in infinite volume.  Results for $F$ in
this $I=2$ case and the quantities used to determine it are given in
Tab.~\ref{tab:LL}.  We should note that applying the finite volume correction
of Eq.~\ref{eq:LL} gives us a finite-volume corrected amplitude for a
$\Delta I = 3/2$, $K\rightarrow \pi\pi$ decay that is slightly above
threshold by the amount $E_2^{\pi\pi}-2m_\pi = 33(1)$ MeV.

\begin{table}
\caption{The calculated quantities which appear in the Lellouch-L\"uscher
factor $F$ for $I=2$.  The corresponding factor for the case of
non-interacting particles is $F_{\rm free}=31.42$.  The difference
reflects the final two-pion scattering in a box.}
\label{tab:LL}
\begin{ruledtabular}
\begin{tabular}{llll}
$p$        & $q\frac{\partial\phi(q)}{\partial q}$
                       & $p\frac{\partial\delta(p)}{\partial p}$
                                      & $F$ \\
\hline
0.0690(13) & 0.221(10) &  -0.0849(43) & 26.01(18) \\
\end{tabular}
\end{ruledtabular}
\end{table}

We can now combine everything and calculate the $K^0\rightarrow\pi\pi$
decay amplitudes,
\begin{equation}
A_{2/0} = F \frac{G_F}{\sqrt{2}}V_{ud}V_{us}\sum_{i=1}^{10}\sum_{j=1}^{7}
          \left[\Bigl(z_i(\mu)+\tau y_i(\mu)\Bigr) 
          Z_{ij}^{{\rm lat}\to{\overline{\rm MS}}} 
          M_j^{\frac{3}{2}/\frac{1}{2},{\rm lat}}\right],
\label{eq:calAI}
\end{equation}
where the construction of the $10 \times 7$ renormalization matrix 
$Z_{ij}^{{\rm lat}\to{\overline{\rm MS}}}$ is explained in Appendix A.
For later use we have written Eq.~\ref{eq:calAI} in a way which is 
applicable for $\Delta I =  1/2$ decays as well as for the $\Delta I=3/2$ 
transitions considered in this section.  The results for the complex 
$\Delta I=3/2$ decay amplitude $A_2$ are summarized in Tab.~\ref{Tab:A2}, 
including those for the other two, energy-non-conserving choices of kaon 
mass.  Since $m_K^{(1)}$ differs from the isospin-2 $\pi-\pi$ energy by only 
0.2 percent, we quote this case as our energy-conserving kaon decay amplitude.
Therefore, in physical units, we obtain the energy-conserving $\Delta I=3/2$,
$K^0\rightarrow\pi\pi$ complex, threshold decay amplitude for $m_K= 877$ MeV
and $m_\pi=422$ MeV:
\begin{eqnarray}
\mathrm{Re}(A_2) & = & 4.911(31)\times 10^{-8} \mathrm{GeV} \\
\mathrm{Im}(A_2) & = & -0.5502(40) \times 10^{-12} \mathrm{GeV}.
\end{eqnarray}
This result for Re$(A_2)$ can be compared with the experimental value
of $1.48\times 10^{-8}$ GeV given above.  The larger result found in our calculation
is likely explained by our unphysically heavy kaon and pions.

\begin{table}
\caption{The complex, $K^0\rightarrow\pi\pi$, $\Delta I=3/2$ decay amplitudes
in units of GeV.}
\label{Tab:A2}
\begin{ruledtabular}
\begin{tabular}{lll}
$m_K$  & Re$(A_2)(\times 10^{-8})$ & Im$(A_2)(\times 10 ^{-12})$ \\
\hline
$m_K^{(0)}$ & 4.308(28) & -0.5596(40) \\
$m_K^{(1)}$ & 4.911(31) & -0.5502(40) \\
$m_K^{(2)}$ & 5.916(38) & -0.5316(39) \\
\end{tabular}
\end{ruledtabular}
\end{table}

\section{$K^0\rightarrow\pi\pi$ $\Delta I=1/2$ amplitude}

\begin{figure*}
\begin{tabular}{cc}
   \includegraphics[width=0.5\textwidth]{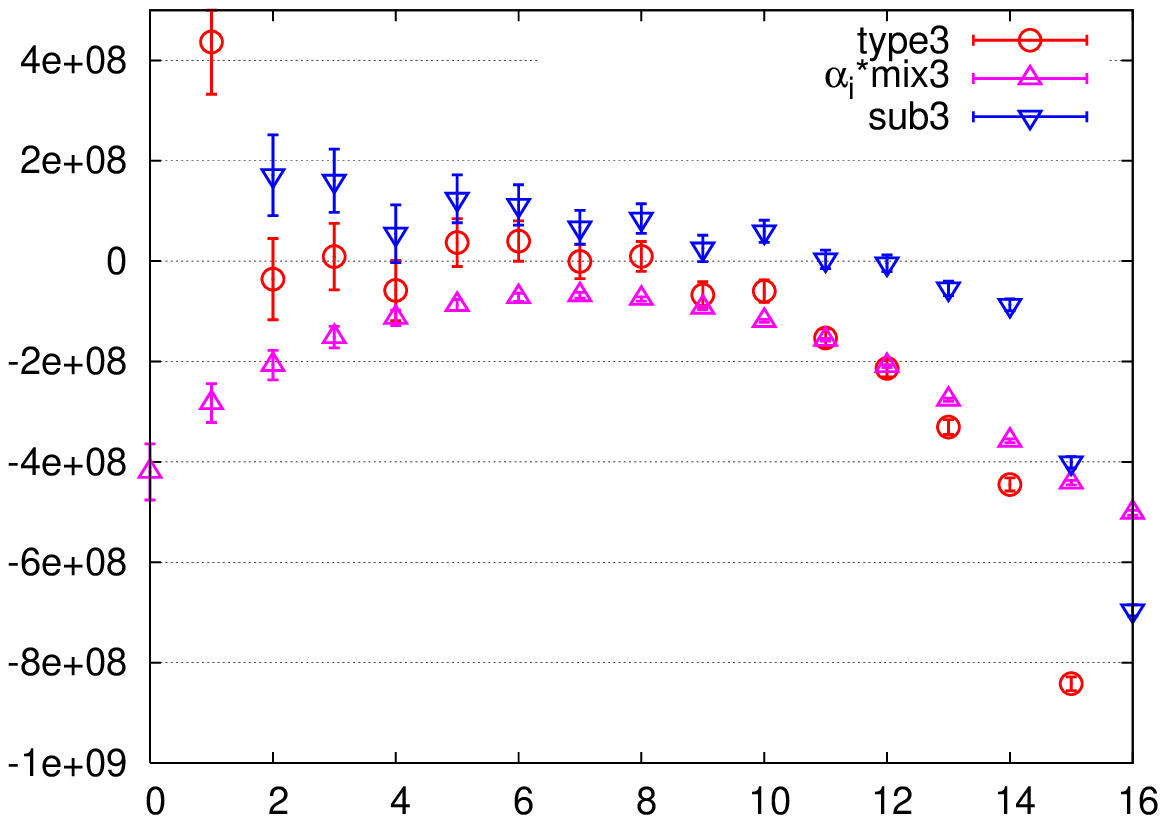} &
   \includegraphics[width=0.5\textwidth]{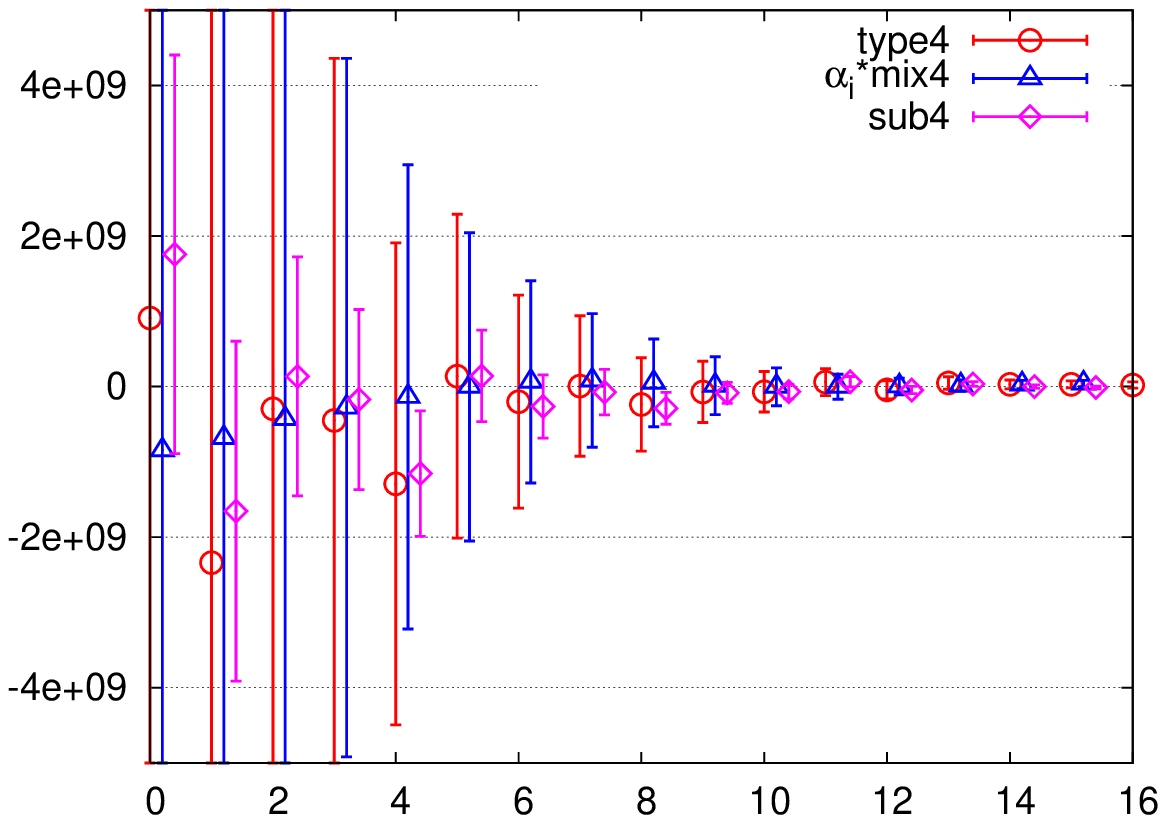} \\
	(a) & (b) \\
   \includegraphics[width=0.5\textwidth]{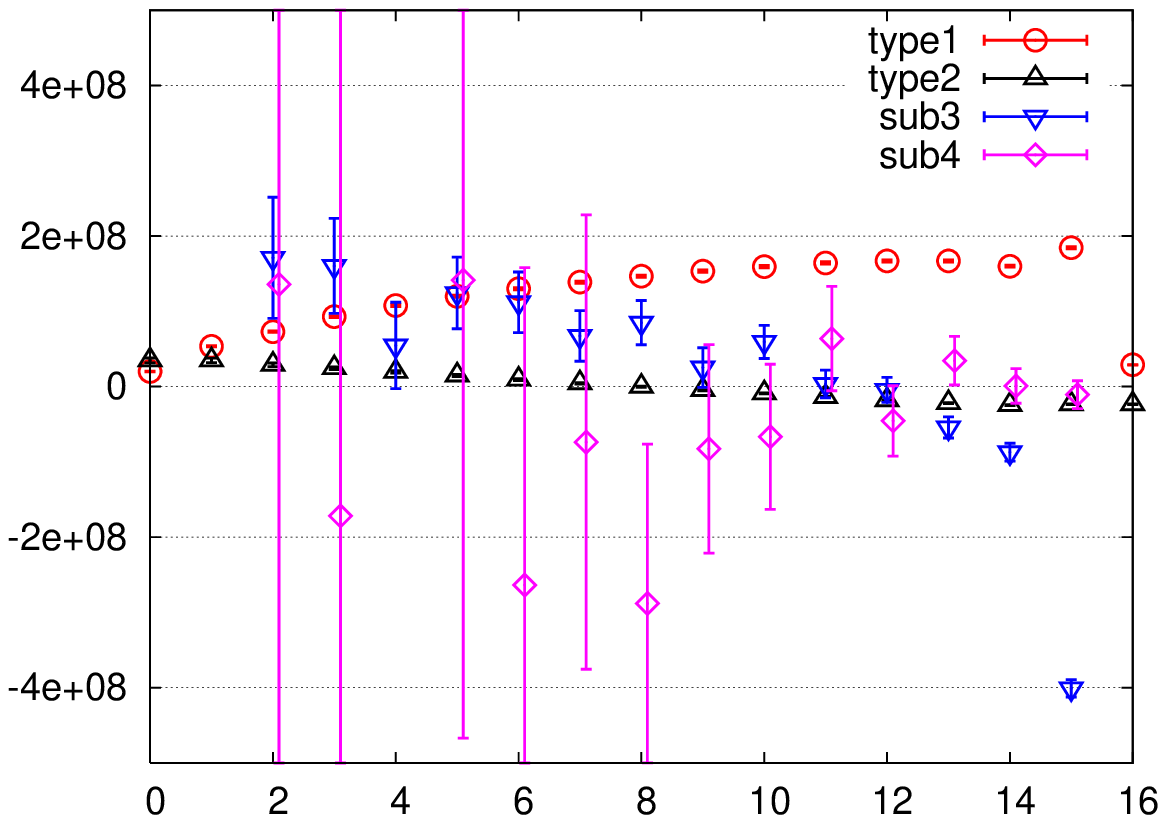} &
   \includegraphics[width=0.5\textwidth]{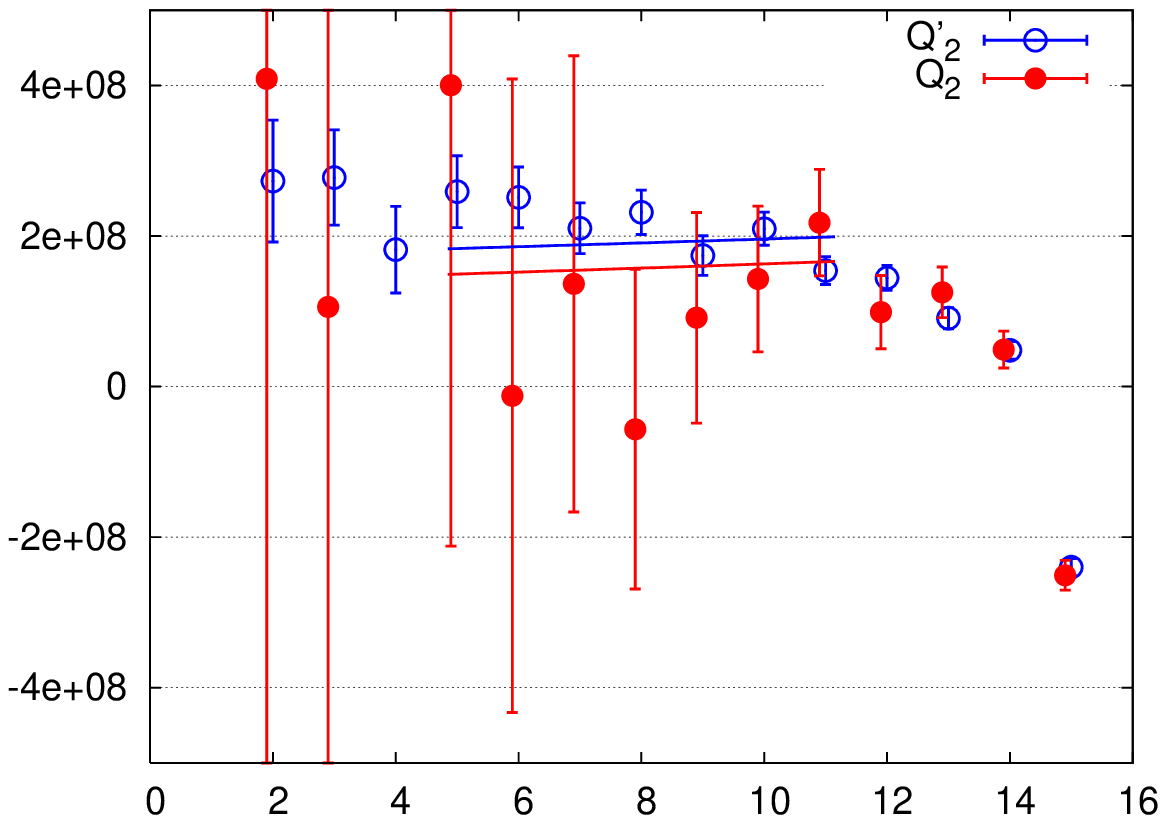} \\
	(c) & (d)
 \end{tabular}
 \caption{Plots showing the $t$ dependence of the various contractions which
contribute to the $\Delta I = 1/2$ correlation function $C_{0,2}(\Delta=16,t)$
for the operator $Q_2$.  (a) Contractions of $type3$, the divergent mixing term
$mix3$ that will be subtracted and the result after subtraction, $sub3$.
(b) Contractions of $type4$, the divergent mixing term $mix4$ that will be
subtracted and the result after subtraction, $sub4$.  (c) Results for each
of the four types of contraction after the needed subtractions have been
performed.  (d): Results for the complete $Q_2$ correlation function
$C_{0,2}(\Delta=16,t)$ obtained by combining these four types of contractions.
The solid points labeled $Q_2$ are the physical result while the open points
labeled $Q_2'$ are obtained by omitting all the vacuum graphs, $sub4$.  The
solid and dotted horizontal lines indicate the corresponding fitting results
and the time interval, $5 \le t \le 11$ over which the fits are performed.}
\label{fig:Q2}
\end{figure*}

\begin{figure*}
\begin{tabular}{cc}
   \includegraphics[width=0.5\textwidth]{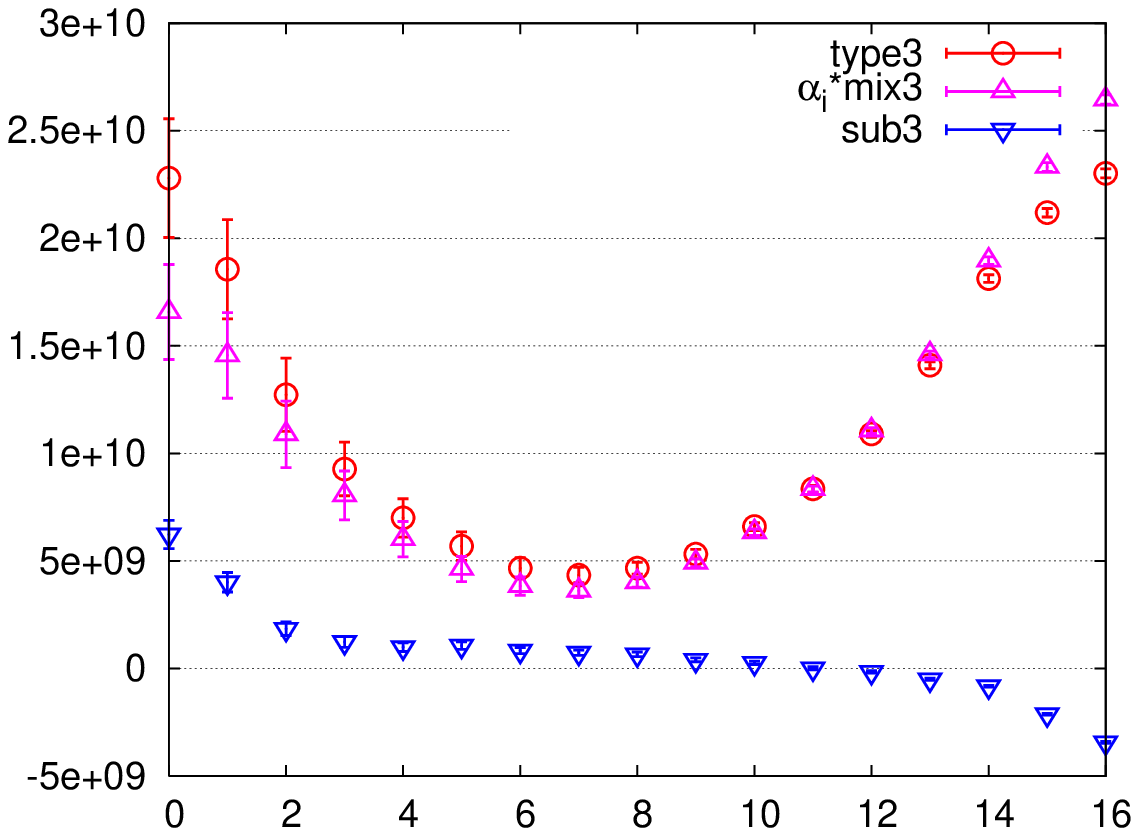} &
    \includegraphics[width=0.5\textwidth]{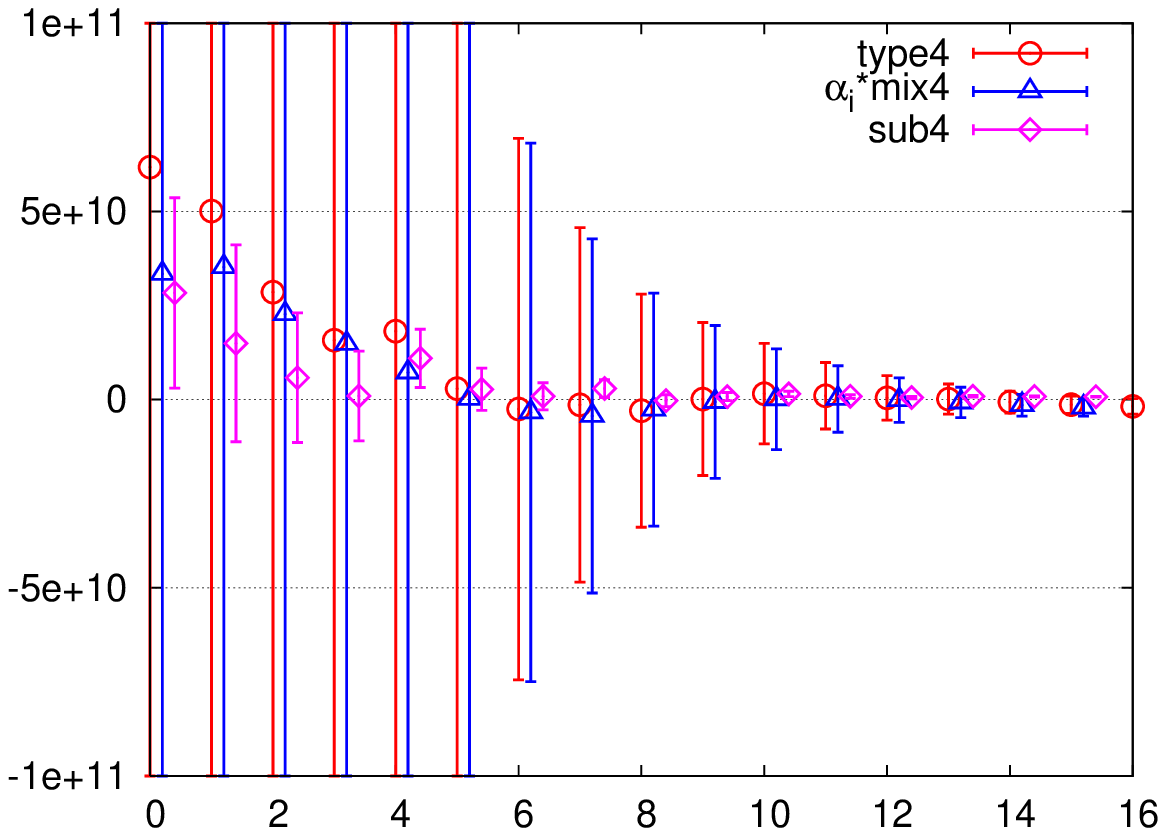} \\
	(a) & (b) \\
   \includegraphics[width=0.5\textwidth]{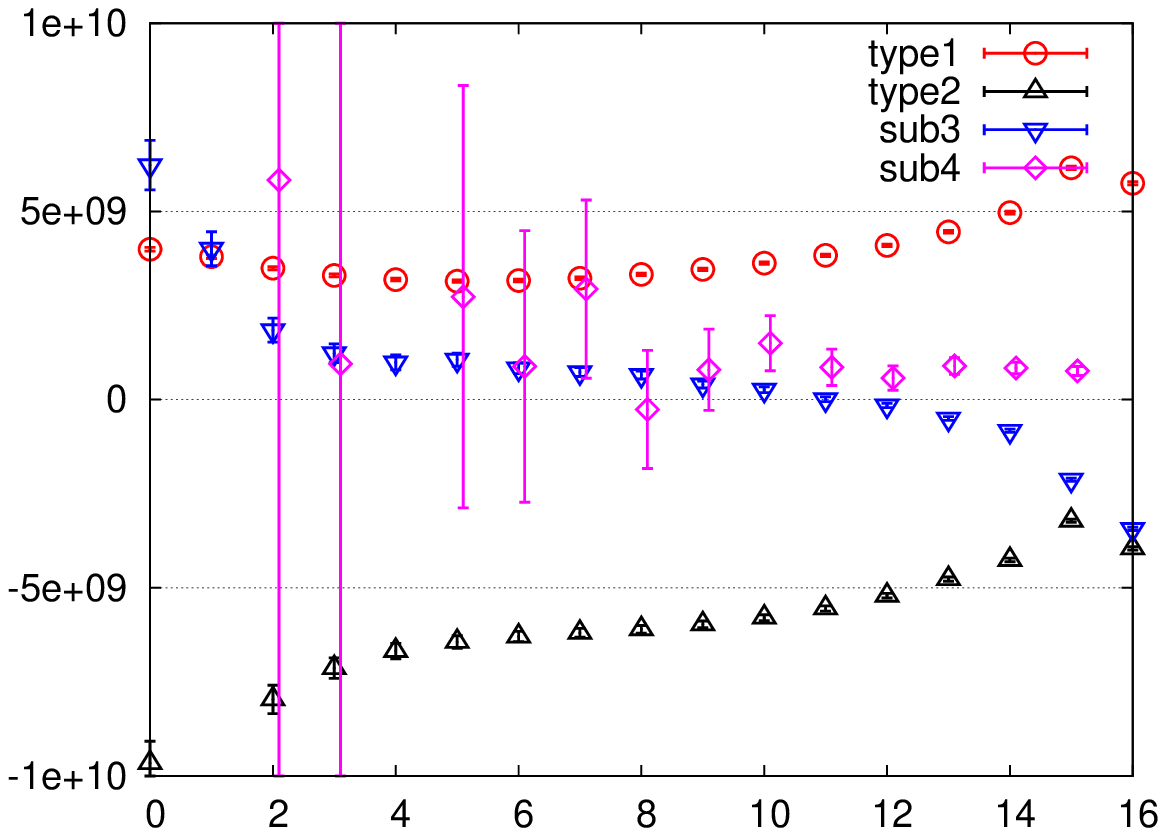} &
    \includegraphics[width=0.5\textwidth]{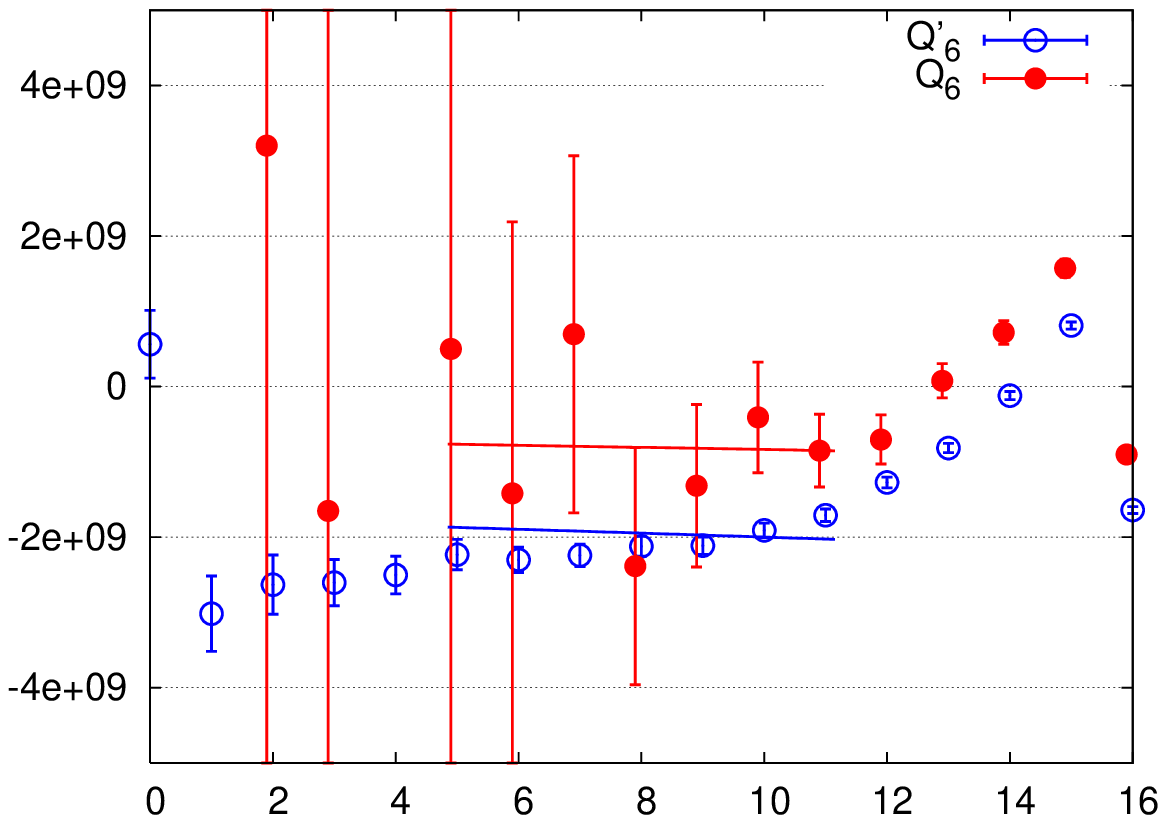} \\
	(c) & (d)
 \end{tabular}
 \caption{The result for each type of contraction contributing to the 3-point
correlation function $C_{0,6}(\Delta=16,t)$ for the operator $Q_6$ following
the same conventions as in Fig.~\ref{fig:Q2}.}
\label{fig:Q6}
\end{figure*}

Following the prescription given by Eq.~\ref{eq:I0} we have calculated all
of the $\Delta I=1/2$ kaon decay correlation functions,
\begin{equation}
C_{0,i}(\Delta, t) =
  \frac{1}{32}\sum_{t'=0}^{31} A_{0,i}(t_\pi=t'+\Delta,t_{\rm op}=t+t',t_K=t'),
\label{eq:I_0_correlator}
\end{equation}
for each of the ten effective weak operators.  In the calculation we treat
each of these ten operators as independent and then verify that the identities
shown in Eq.~\ref{eq:identity} are automatically satisfied.  Figures~\ref{fig:Q2}
and \ref{fig:Q6} show two examples of the resulting correlation functions
for the operators $Q_2$ and $Q_6$, in the case of the lightest kaon $m_K^{(0)}$.
Table~\ref{tab:mass} shows that the mass of this kaon is very close to the
energy of the I=0 two-pion state.  Therefore, we expect to get a reasonably
flat plateau when the operator is far from both the source and sink.

Given this good agreement between the energies of the $K$ and
$\pi-\pi$ states, we might expect that the unphysical, dimension three
operator, $\overline{s}\gamma^5d$ which mixes with the $(8,1)$ operators
in Eq.~\ref{eq:weak_eff} and is itself a total divergence, will also give
a negligible contribution to such an energy and momentum conserving matrix
element.  However, as can be seen from Figs.~\ref{fig:Q2}(a) and
\ref{fig:Q6}(a), the matrix element of this term is large and the explicit
subtraction described in Sec.~\ref{sec:K_decay_contractions} is necessary.

This difficulty is created by the combination of two phenomena.  First
the mixing coefficient which multiplies the $\overline{s}\gamma^5d$
operator when it appears in our weak $(8,1)$ operators is large, of
order $(m_s-m_l)/a^2$.  Second, in our lattice calculation the necessary
energy conserving kinematics (needed to insure that this total divergence
does not contribute) is only approximately valid.  The required equality of
the spatial momenta of the kaon and $\pi - \pi$ states is assured by our
summing the location of the weak vertex over a complete temporal hyperplane.
On the other hand, the equality of the energies of the initial and final 
states results only if we have adjusted the kaon mass to approximately 
that of the two-pion state and chosen the time extents sufficiently large 
that other states with different energies have been suppressed.
However, as can be seen in Figs.~\ref{fig:Q2}(a) and \ref{fig:Q6}(a) the
subtraction terms $mix$3 and $mix$4 show strong dependence on the time at
which they are evaluated.  This implies that there are important contributions
coming from initial and final states which have significantly different
energies.  One or both of these states is then not the intended $K$ or
$\pi-\pi$ state but instead an unwanted contribution which has been
insufficiently suppressed by the time separations between source, weak
operator and sink.

Thus, instead of relying on large time extents and energy conserving
kinematics to suppress this unphysical, $O(1/a^2)$ term we must explicitly
remove it.  As explained in Sec.~\ref{sec:K_decay_contractions} this can
be done by including an explicit subtraction which we fix by the requirement
that the kaon to vacuum matrix element of the complete subtracted operator
vanishes as in Eq.~\ref{eq:subtraction_condition}.  Thus, we determine the
divergent coefficient of this mixing term from the ratio
$\alpha_i = \langle 0|Q_i|K \rangle/\langle 0|\overline{s}\gamma^5d|K \rangle$
and then perform the explicit subtraction of the resulting terms,
labeled $\alpha_i\cdot mix3$ and $\alpha_i\cdot mix4$ in Figs.~\ref{fig:Q2}
and \ref{fig:Q6}.

Of course, the finite part of such a subtraction is not determined from first
principles and our choice, specified by Eq.~\ref{eq:subtraction_condition}
is arbitrary.  Thus, we must rely on our identification of a plateau and
the approximate energy conservation of our kinematics to make the arbitrary
part of this subtraction small, along with the other errors associated with
evaluating the decay matrix element of interest between initial and final
states with slightly different energies.

We now examine the very visible time dependence in Figs.~\ref{fig:Q2}(a)
and \ref{fig:Q6}(a) for both the original matrix elements and the
subtraction terms in greater detail.  As discussed above one might
expect these divergent subtraction terms to contribute to excited state
matrix elements in which the energies of the initial and final states
are very different.  Typical terms should be exponentially suppressed
as the separation between the weak operator and the source or sink is
increased, with the time behavior $\exp\{-(m_K^*-m_K)t\}$ or
$\exp\{-(E_{\pi\pi}^*-E_{\pi\pi})(\Delta -t)\}$, which ever is larger. 
(The $\ast$ denotes an excited state.) However, by carefully examining 
the time behavior of the $mix3$ amplitude,
we find that the time dependence, at least in the vicinity of the central
region, is less rapid than might be expected from such excited states
suggesting that it is probably not due primarily to contamination
from excited states.

\begin{figure}
\begin{tabular}{c}
\includegraphics[width=0.4\textwidth]{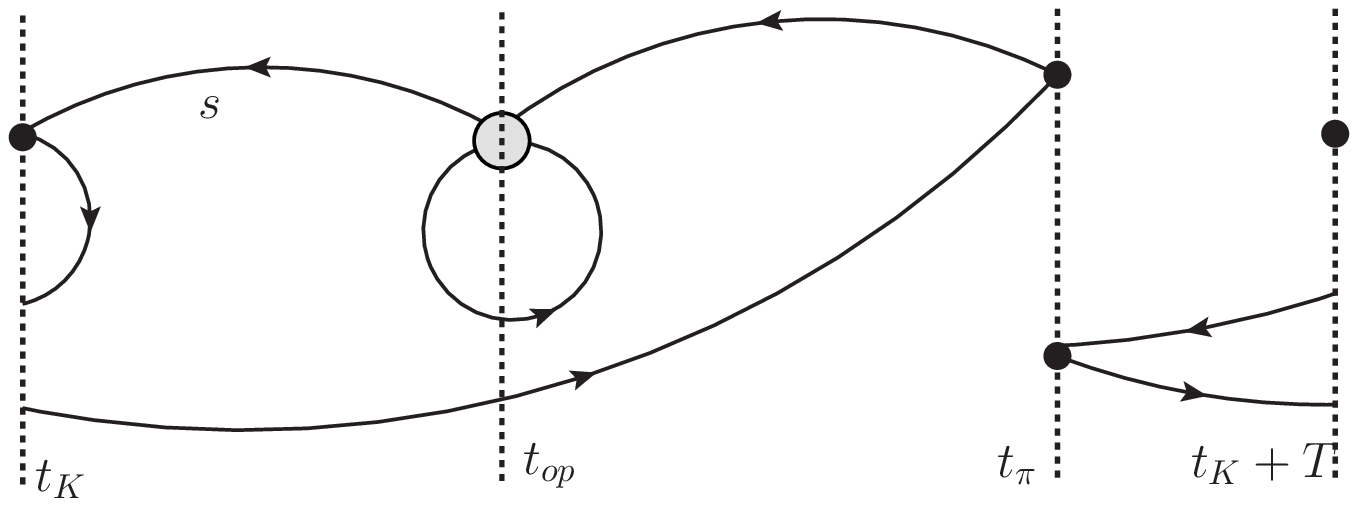} \\
\includegraphics[width=0.4\textwidth]{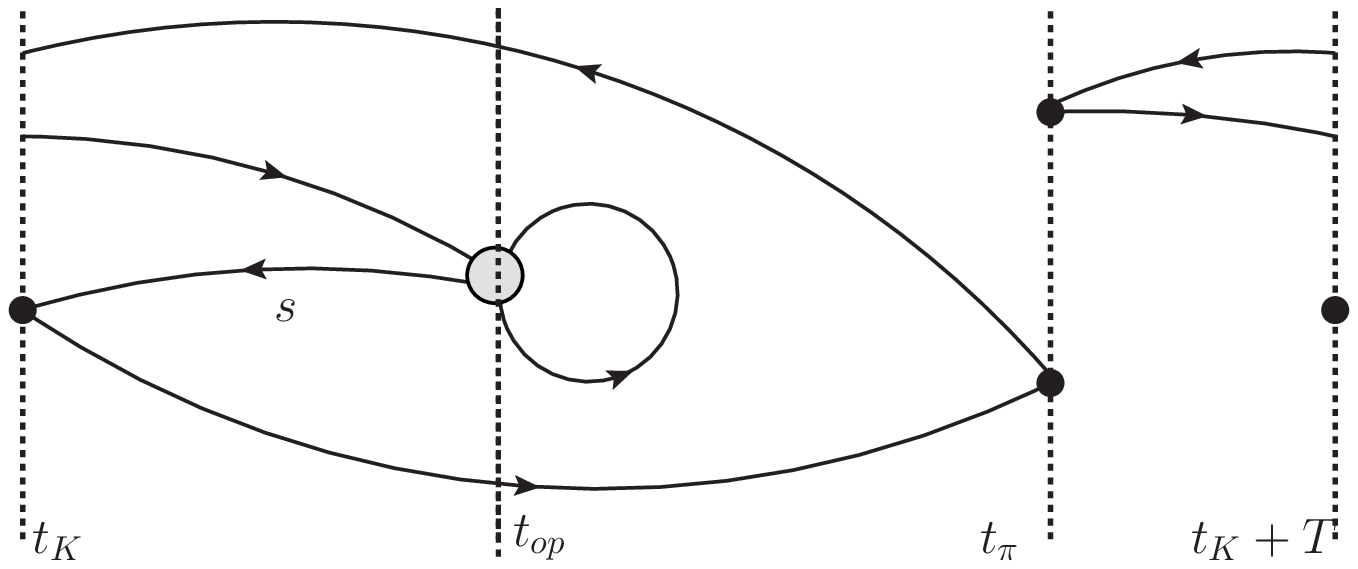} \\
\includegraphics[width=0.4\textwidth]{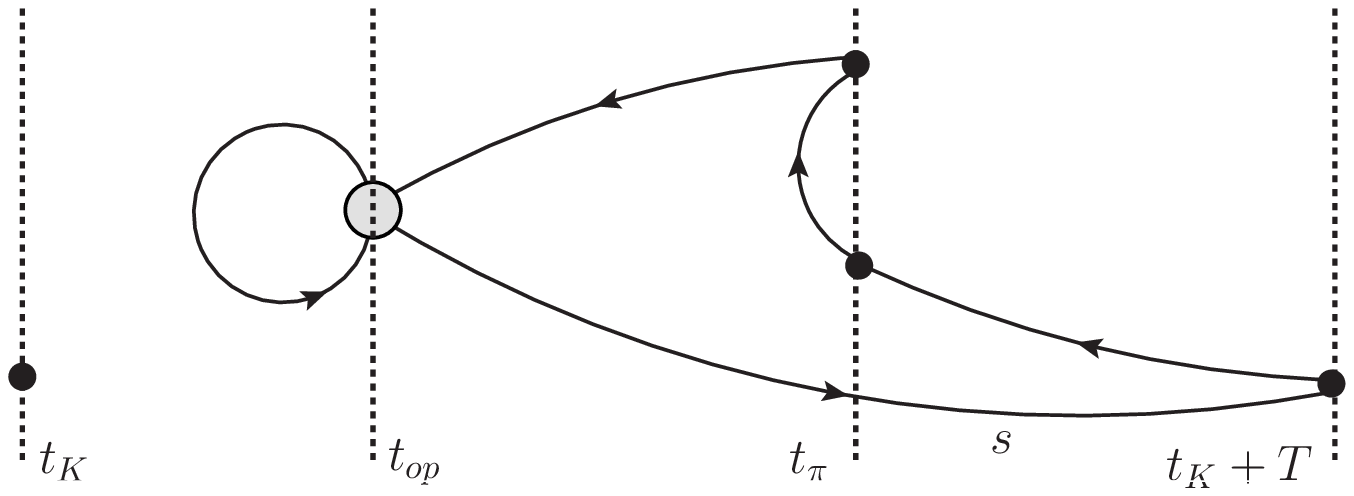}
\end{tabular}
\caption{The dominant around-the-world paths contributing to graphs of
$type3$.  As in Fig.~\protect{\ref{fig:type1round}} we show the space-time
region between the kaon source at $t=t_K$ and its periodic recurrence at
$t=t_K+T$.  The gray circle represents the four quark operator $Q_i$.  For
the first two graphs, one of the two pions created at the $t=t_\pi$ source
travels directly to the operator $Q_i$ while the second pion travels in the
other direction in time and reaches the kaon and weak operator by passing
through the periodic lattice boundary.  In the third diagram it is the kaon
which travels in the opposite to the expected time direction.}
\label{fig:type3round}
\end{figure}

We believe that the dominant, energy-nonconserving matrix
elements which cause the significant time dependence in Figs.~\ref{fig:Q2}
and \ref{fig:Q6} arise from the around-the-world effects identified
and discussed in the previous $\Delta I = 3/2$ section.  In fact,
for the reasons just discussed associated with divergent operator
mixing, such around-the-world effects are a more serious problem in
the $\Delta I=1/2$ case.   The dominant around-the-world graphs are
shown in Fig.~\ref{fig:type3round}.  An estimate of the time 
dependence of these graphs gives,
\begin{eqnarray}
&&<K^0 \pi|Q_i|\pi>N_\pi N_K N_\pi e^{-m_\pi T}  e^{-(E_{K\pi}-m_\pi)t} 
\nonumber \\ &&
+ <0|Q_i|K^0\pi\pi>N_\pi N_K N_\pi e^{-m_K((T-\Delta)+(\Delta-t))}\,,
\end{eqnarray}
where the first term comes from the first two graphs of
Fig.~\ref{fig:type3round}, while the second term comes from the third
graph. (Recall that $t=t_\mathrm{op}-t_K$ and $\Delta = t_\pi-t_K$).
Notice that these two terms involve amplitudes which are far from
energy conserving and therefore contain large divergent contributions
from mixing with the operator $\overline{s}\gamma_5 d$ which will be
removed only when combined with the corresponding around-the-world
paths occuring in the $mix3$ contraction.

We conclude that it is these around-the-world matrix elements
which are the reason for the observed large divergent subtraction 
in the $type3$ graph.  The largest divergent contribution is thus not 
the subtraction for the matrix element we are trying to evaluate,
$<\pi\pi|Q_i|K^0>$; rather, it is the divergent subtraction for
the matrix elements $<K^0\pi|Q_i|\pi>$ and $<0|Q_i|K^0\pi\pi>$ which
arise from the around-the-world paths which are not sufficiently
suppressed by our lattice size.  Two important lessons can be
learned from this analysis.  First, it is important to perform
an explicit subtraction of the divergent mixing with the operator
$\overline{s}\gamma_5 d$.  While this term will not contribute to
the energy conserving matrix element of interest, in a Euclidean
space lattice calculation there are in general, other, unwanted, 
energy non-conserving terms which may be uncomfortably large if
this subtraction is not performed.  Second it would be wise to work 
on a lattice with a much larger size $T$ in time direction in order 
to suppress further the around-the-world terms which give such a
large contribution in the present calculation. Using the average
of propagators computed with periodic plus anti-periodic boundary
conditions to effectively double the length in the time direction
would be a good solution.

We should emphasize that these divergent, around-the-world
contributions do not pose a fundamental difficulty.  The largest
part of these amplitudes are removed by the corresponding subtraction
terms constructed from the operator $\overline{s}\gamma_5d$.  The
remaining finite contributions from this and other around-the-world
terms are suppressed by the factor $\exp(-m_\pi T)$ or
$\exp(-m_k(T-\Delta))$.  Fortunately, the large divergent subtraction
also reduces the statistical errors substantially, especially for the
$type4$ vacuum graphs, which indicates the expected strong correlation
between the divergent part of the weak operator and the corresponding
$\overline{s}\gamma_5d$ subtraction.  Our results suggest that the
separation of $\Delta=16$ gives a relatively longer plateau region,
so we use that $K-\pi\pi$ time separation in the analysis below.

The lattice matrix elements are determined by fitting the $I=1/2$ correlators
$C_0^i(\Delta, t)$ given in Eq.~\ref{eq:I_0_correlator} using the fitting form:
\begin{eqnarray}
C_{0,i}(\Delta, t)& = & M_i^{1/2,{\rm lat}} N_{\pi\pi} N_K
                   e^{-E_{\pi\pi}\Delta} e^{-(m_K-E_{\pi\pi})t}.
\label{eq:I_0_fit}
\end{eqnarray}
The fitted results for the weak, $\Delta I=1/2$ matrix elements of all ten
operators are summarized in Tab.~\ref{Tab:Qi}.  To see the effects of the
disconnected graph clearly, a second fit is performed to the amplitude from
which the disconnected, $type4$ graphs have been omitted and the calculated
results are shown with an additional $\prime$ label, as in the earlier two-pion
scattering section.

\begin{table*}
\caption{Fitted results for the weak, $\Delta I=1/2$ kaon decay matrix
elements using the kaon mass $m_K^{(0)}$.  The column $M_i^{\rm lat}$
shows the complete result from each operator. The column
$M_i^{\prime\,{\rm lat}}$ shows the result when the disconnected graphs
are omitted while the 4th and 5th columns show the contributions of each
operator the real and imaginary parts of the physical decay amplitude $A_0$.
These results are obtained using a source-sink separation $\Delta = 16$,
and  a fit range $5 \le t \le  11$.}
\label{Tab:Qi}
\begin{ruledtabular}
\begin{tabular}{lllll|}
i  & $M_i^{1/2,{\rm lat}}(\times 10^{-2})$
               & $M_i^{\prime 1/2,{\rm lat}}(\times 10^{-2})$
			   &  Re$(A_0)$(GeV) 
			                        & Im$(A_0)$(GeV)\\
\hline
1  & -1.6(16) & -1.10(37)  & 7.6(64)e-08     &    0            \\
2  & 1.52(61) & 1.92(15)   & 2.86(97)e-07     &    0            \\
3  & -0.3(41) & 0.3(10)    & 2.1(136)e-10   &    1.1(76)e-12  \\
4  & 2.7(33)  & 3.32(78)   & 4.2(44)e-09     &    1.4(14)e-11  \\
5  & -3.3(38) & -6.81(86)  & 3.1(53)e-10    &    1.6(28)e-12  \\
6  & -7.8(48) & -19.6(9)   & -5.6(33)e-09   &    -3.3(20)e-11 \\
7  & 10.9(14) & 15.20(42)  & 5.2(12)e-11     &    8.8(20)e-14  \\
8  & 35.7(28)&  47.2(10)  & -3.66(28)e-10   &    -1.79(14)e-12\\
9  & -2.2(12) & -1.79(29)  & 3.1(15)e-14   &    -2.01(96)e-12 \\
10 & 0.9(12)  & 1.24(29)   & 1.2(11)e-11     &    -2.7(27)e-13 \\

\hline                     	                        
Total & - & - & 3.46(78)e-07 & -2.4(23)e-11 \\
\end{tabular}
\end{ruledtabular}
\end{table*}

The calculation of the $\Delta I = 1/2$ decay amplitude $A_0$ from the
lattice matrix elements $M_i^{1/2,{\rm lat}}$ given in Tab.~\ref{Tab:Qi} is
very similar to the $\Delta I=3/2$ case: the values of $M_i^{1/2,{\rm lat}}$
are simply substituted in Eq.~\ref{eq:calAI}.  However, the attractive
character of the $I=0$, $\pi-\pi$ interaction and resulting negative value of 
$p^2$ makes the Lellouch-L\"uscher treatment of finite volume corrections
inapplicable.  For the repulsive $I=2$ case, we could apply this treatment
to obtain the decay amplitude for a two-pion final state which was slightly
above threshold corresponding to the actual finite volume kinematics.
In the present case there is no corresponding infinite-volume decay into
two pions below threshold and an unphysical increase of $m_\pi$ to compensate
for the finite volume $\pi-\pi$ attraction will introduce an $O(1/L^3)$ error
in the decay amplitude of the same size as that which the Lellouch-L\"uscher
treatment corrects.  Thus, for this $\Delta I=1/2$ we do not include finite
volume corrections and simply use the free-field value for the factor $F$
in Eq.~\ref{eq:calAI}.

While we believe that we cannot consistently apply the Lellouch-L\"uscher
finite volume correction factor to improve our result for the $I=0$,
$K \rightarrow \pi\pi$ decay amplitude, we might still be able to use
the quantization condition of Eq.~\ref{eq:FV_quant} to determine the
$I=0$ $\pi-\pi$ scattering phase shift $\delta_0(p)$.  Even though
Eq.~\ref{eq:FV_quant} can be analytically continued to imaginary values
of the momentum $p$, its application for large negative $p^2$ is uncertain
since the function $\phi(q)$ becomes ill defined.  In fact, our value of
$p^2$ sits very close to a singular point of $\phi(q)$.   We believe this
happens because the condition on the interaction range $R \ll L/2$ used to
derive the quantization condition in Eq.~\ref{eq:FV_quant} is not well
satisfied for our small volume.  This impediment to determining $\delta_0(p)$
will naturally disappear once we work with lighter pions in a larger volume.

The results for Re($A_0$) and Im($A_0$) are summarized in Tab.~\ref{Tab:A0}
and the individual contribution from each of the operators is detailed in
the last two columns of Tab.~\ref{Tab:Qi}.  Within a large uncertainty
Tab.~\ref{Tab:Qi} shows that the largest contribution to Re($A_0$) comes
from operator $Q_2$, and that to Im($A_0$) from $Q_6$ as found, for example,
in Refs.~\cite{Blum:2001xb,Noaki:2001un}.

Since the choice $m_K^{(0)}$ for the kaon mass is not precisely equal to
the energy of the $I=0$ $\pi\pi$ state, we carried out a simple linear
interpolation between $m_K^{(0)}$ and $m_K^{(1)}$ to obtain an energy conserving
matrix element, which is shown in the last row of Tab~\ref{Tab:A0}.  In terms of
physical units, therefore, our full calculation gives the energy conserving,
$K^0 \rightarrow \pi\pi$, $\Delta I=1/2$, complex decay amplitude $A_0$ for
$m_K=766$ MeV and $m_\pi=422$ MeV:
\begin{eqnarray}
\mathrm{Re}(A_0) & = &  3.80(82)\times10^{-7}  \mathrm{GeV}\\
\mathrm{Im}(A_0) & = &  -2.5(2.2)\times10^{-11} \mathrm{GeV}.
\end{eqnarray}
These complete results can be compared with those obtained when the
disconnected graphs are neglected given in Tab.~\ref{Tab:A0} and
the experimental value for Re$(A_0)=3.3 \times 10^{-7}$ GeV.  As in the case
of Re$(A_2)$, our larger value is likely the result of our unphysically
heavy kaon and pion.

\begin{table*}
\caption{Amplitudes for $\Delta I=1/2$ $K^0 \rightarrow \pi\pi$ decay in
units of GeV.  The energy conserving amplitudes are obtained by a simple
linear interpolation between $m_K^{(0)}$=0.42599 and $m_K^{(1)}$=0.50729
to the energy of two-pion state.  As in the previous tables, the $\prime$
indicates results from which the disconnected graphs have been omitted.}
\label{Tab:A0}
\begin{ruledtabular} 
\begin{tabular}{lllllll}
$m_K$             & Re$(A_0)(\times 10^{-8})$
                                & Re$(A^\prime_0)(\times 10^{-8})$ 
                                              & Im$(A_0)(\times 10^{-12})$ 
                                                          & Im$(A^\prime_0)(\times 10 ^{-12})$ \\
\hline 
$m_K(0)$          & 36.1(78)  & 42.3(20)  & -21(21) & -66.1(43)  \\
$m_K(1)$          & 45(10)    & 48.8(24)  & -41(26) & -74.6(47)  \\
$m_K(2)$          & 65(15)    & 58.6(32)  & -69(39) & -89.6(63) \\
Energy conserving & 38.0(82)  & 43.4(21)  & -25(22) & -67.5(44)  \\
\end{tabular} 
\end{ruledtabular}
\end{table*}

\section{Discussion and Conclusions}

Comparing the results of Re($A_2$) in Tab.~\ref{Tab:A2} and Re($A_0$) in
Tab.~\ref{Tab:A0}, we find the $\Delta I=1/2$ enhancement ratio
Re($A_0$)/Re($A_2$) to be roughly 7-9.  This comparison is degraded by our
threshold kinematics which, since the $I=0$ and $I=2$ two-pion states have 
different energies in a finite volume, causes us to use a different kaon 
mass in the calculations of $(A_2)$ and $(A_0)$ in order to have energy 
conserving decays in each case.  These two energy conserving amplitudes 
have a ratio of $38.0/4.911=7.7$, while if we ignore energy conservation 
and use the same $m_K^{(1)}$ value for kaon mass, the ratio becomes 
$45.0/4.911=9.2$.  Of course, both estimates are far from the experimental 
ratio of 22.5 suggesting that our 422 MeV pion mass and small lattice 
volume are far from physical.

For completeness, we also calculate the measure of direct CP violation,
\begin{equation}
\mathrm{Re}\left(\frac{\epsilon^\prime}{\epsilon}\right)
   =\frac{\omega}{\sqrt{2}|\epsilon|}\left[\frac{\mathrm{Im}(A_2)}{\mathrm{Re}(A_2)}
                  -\frac{\mathrm{Im}(A_0)}{\mathrm{Re}(A_0)}\right],
\end{equation}
where $\omega=\mathrm{Re}(A_2)/\mathrm{Re}(A_0)$ is the inverse of the
$\Delta I=1/2$ enhancement factor.  Using our kinematics, the kaon mass
$m_K^{(1)}$ and substituting the experimental value for $\epsilon$, we get
Re$(\epsilon^\prime/\epsilon)= (2.7 \pm 2.6)\times10^{-3}$.  If we instead use
the experimental value for $\omega$, we get
Re$(\epsilon^\prime/\epsilon)= (1.11 \pm 0.91)\times10^{-3}$.

Our calculation is sufficiently far from physical kinematics, that it is
not appropriate to compare these results with experiment.\footnote{A further
unphysical aspect of our kinematics is the inequality of the strange quark 
mass used in the fermion determinant and the self contractions appearing in
the eye graphs ($m_s=0.032$) and strange quark masses used in the valence 
propagator of the K meson ($m_s=0.066$, 0.99 and 0.165).} Instead, our
objective is to show how well our method performs.  We have been able to
calculate Re($A_0$), the key element needed to explain the $\Delta I=1/2$
rule, with a 25\% statistical error.  Comparing our results for Re($A_0$)
obtained on sub-samples of N=100, 400 and all 800 configurations we
find that the statistical errors on the quantities we measure do indeed scale
as $1/\sqrt{N}$.   Therefore, we believe that our non-zero signal for
Re($A_0$) is real and that we could reduce this statistical error to
10 percent by quadrupling the size of our sample to 3200 configurations.
It is interesting to note the results for primed (disconnected graphs
omitted) and unprimed (all graphs included) quantities contributing to
Re$(A_0)$ have similar values suggesting that the disconnected graphs,
while contributing significantly to the statistical error, have an
effect on the final result for Re$(A_0)$ at or below 25\%.

In contrast, the result for Im($A_0$) has an 80\% error.  Thus, it is not
clear whether the size of the result will survive a quadrupling of the
sample with its statistical error reducing to a 40\% error or whether the
result itself will shrink, remaining statistically consistent with zero.
Considering the substantial systematic errors associated with our small
volume and the fact that our kinematics are far from the physical, we
present this trial calculation as a guideline for future work and a proof
of method rather than giving accurate numbers to compare with experiment.

From our observation of the around-the-world effect, we conclude
that it is important to use the average of quark propagators obeying
periodic and anti-periodic boundary conditions to extend the lattice
size in the time direction.   In addition, explicit subtraction of
the divergent mixing term $\overline{s}\gamma^5 d$ is necessary
even for kinematics which are literally energy conserving because
the around-the-world path and possibly other excited state matrix
elements are far off shell and can be substantially enhanced by such
a divergent contribution.  Finally, future work should be done using a
much larger lattice which can contain two pions without any worry about
finite size effects.

The focus of this paper is on developing techniques capable
of yielding statistically meaningful results from the challenging
lattice correlation functions involved in the amplitude $A_0$.
However, there are other important problems that will also require
careful attention if physically meaningful results are to be
obtained for this amplitude with an accuracy of better than 20\%.
Two important issues are associated with operator mixing.  As
discussed in Appendix A, a proper treatment of the non-perturbative
renormalization of the four independent $(8,1)$ four-quark operators
requires that additional operators containing gluonic variables
(some of which are not gauge invariant) be included.  While including
such operators is in principle possible and the subject of active
research, controlling such mixing using RI/MOM methods offers
significant challenges.

A second problem is operator mixing induced by the residual chiral
symmetry breaking of the DWF formulation.  The mixing of such
wrong-chirality operators should be suppressed by a factor of
order $m_{\rm res}$.  However, the $K \rightarrow \pi\pi$ matrix
elements of the important $(8,1)$ four-quark operators are themselves
suppressed by at least one power of $m_K^2$, a suppression that
is absent from similar matrix elements of the induced, wrong-chirality
operators.  Therefore, such mixing has been ignored in this paper
because its effect on the matrix elements of interest are expected
to be of order $m_{\rm res}/m_s \approx 0.08$, suggesting that these
effects will be smaller than our 25\% statistical errors.  To
perform a more accurate calculation in the future, these mixing
effects may be further suppressed by adopting a gauge action with
smaller residual chiral symmetry breaking.  For example, this ratio
reduces to 0.04 for the DSDR gauge action now being used in RBC/UKQCD
simulations~\cite{Renfrew:2009wu} and to 0.023 for those ensembles
with the smallest lattice spacing created to date using the Iwasaki
gauge action~\cite{Aoki:2010dy}.  When greater accuracy is required
either an improved fermion action, larger $L_s$ or explicit subtraction
of wrong-chirality mixing must be employed.

As we move closer to the physical pion mass we must overcome a
further important difficulty: giving physical relative momentum to
the two pions.  This can be accomplished while keeping the two-pion
state in which we are interested as the ground state, if the kaon
is given non-zero spatial momentum relative to the lattice.  In
this case the lowest energy final state can be arranged to have one
pion at rest while the other pion carries the kaon momentum, as in
the $\Delta I=3/2$ calculation of Ref.~\cite{Yamazaki:2008hg}.
However, this requires the momentum carried by the initial kaon and
final pion to be 739 MeV, which is 5.4 times larger than the physical
pion mass.  Such a large spatial momentum will likely make the
calculation extremely noisy.  For the $\Delta I=3/2$ calculation,
it is possible to use anti-periodic boundary conditions in one or
more spatial directions for one of the light quarks so that each
pion necessarily carries the physical, 206\,MeV momentum present in the
actual decay while the kaon can be at rest~\cite{Kim:2003xt,
Goode:2011kb}.  However, this approach cannot be used in the
case of the $I=0$ final state being studied here.  Instead, the
use of G-parity boundary conditions~\cite{Kim:2002np} may be the
solution to this problem.

\begin{acknowledgments}
We thank Dirk Br\"ommel and our other colleagues in 
the RBC and UKQCD collaborations for discussions, suggestions, and 
assistance.  We acknowledge RIKEN BNL Research Center, the Brookhaven 
National Laboratory and the U.S. Department of Energy (DOE) for providing 
the facilities on which this work was performed.   NC, QL, RM were
supported in part by U.S.~DOE grant DE-FG02-92ER40699, TB and RZ by
U.S.~DOE grant DE-FG02-92ER40716 and  AS and TI by DOE contract
DE-AC02-98CH10886(BNL). EG was supported by an STFC studentship 
and CTS was partially supported by UK STFC Grant PP/D000211/1 
and by EU contract MRTN-CT-2006-035482 (Flavianet).  Finally, 
QL would like to thank the U.S.~DOE for support as a DOE Fellow 
in High Energy Theory and CL acknowledges support of the RIKEN 
FPR program.
\end{acknowledgments}

\appendix

\section{Operator normalization}

In order to combine our lattice matrix elements with the Wilson 
coefficients describing the short-distance weak interaction physics 
responsible for $K \rightarrow \pi\pi$ decay we must 
convert our lattice operators into those normalized according to 
that $\overline{\mathrm{MS}}$ scheme in which the Wilson coefficients 
are evaluated.  We will discuss the details of this procedure in 
this appendix.

The first step is converting the lattice operators into those normalized
according to the RI/MOM scheme~\cite{Martinelli:1995ty}.  We follow the
procedure of Ref.~\cite{Blum:2001xb} and make use of the fact that the
ten operators which enter the conventional expression given in 
Eq.~\ref{eq:weak_eff} are linearly dependent and can be reduced to a set
of seven independent operators, $Q_1'$, $Q_2'$, $Q_3'$, $Q_5'$, $Q_6'$, 
$Q_7'$ and $Q_8'$ defined in Eq.~172-175 Ref.~\cite{Blum:2001xb}.  
These have been defined so that the resulting operators belong to 
specific irreducible representations of $SU_L(3)\times SU_R(3)$.
The operator $Q_1'$ transforms as a $(27,1)$.  The four operators $Q_2'$, 
$Q_3'$, $Q_5'$ and $Q_6'$ all belong to the $(8,1)$ representation, while 
$Q_7'$ and $Q_8'$ each transform as an $(8,8)$.  Here $(m,n)$ denotes 
the product of an $m$-dimensional irreducible representation of $SU_L(3)$ 
with an $n$-dimensional irreducible representation of $SU_R(3)$.  We 
refer to the basis of these seven independent operators as the 
chiral basis.  Because $SU_L(3)\times SU_R(3)$ is an exact symmetry 
of the large momentum, massless limit which our NPR calculation is 
intended to approximate, the mixing matrix $Z^{{\rm lat}\to{\rm RI}}$ given in 
Eq.~\ref{eq:lat2RI} which relates the lattice and RI-normalized operators 
will be block diagonal, only connecting operators which belong to the 
same irreducible representation of $SU_L(3)\times SU_R(3)$.

The RI/MOM conditions which define the operators $O^\mathrm{RI}_i$ and
determine the $7 \times 7$ matrix $Z^{{\rm lat}\to{\rm RI}}$ are imposed 
on the Green's functions:\footnote{While this equation agrees with Eqs.~143 
and 152 of Ref.~\protect{\cite{Blum:2001xb}}, a different choice of momenta
was actually used in that earlier reference.  These two equations accurately
describe the earlier kinematics only after one pair of the momenta $p_1$ and 
$p_2$ are exchanged: $p_1\leftrightarrow p_2$.}
\begin{equation}
G_i(p_1,p_2)_{\alpha\beta\gamma\delta}^f = \prod_{i=1}^4\left\{\int d^4  x_i\right\} 
               \left\langle s(x_1)_\alpha f(x_2)_\beta Q^\mathrm{RI}_i(0)
               \overline{d}_\gamma(x_3) \overline{f}_\delta(x_4)\right\rangle
                e^{-ip_2(x_1+x_2)} e^{ip_1(x_3+x_4)}
\label{eq:2010_BK_kinematics}
\end{equation}
evaluated for $p_1^2 = p_2^2 = (p_1-p_2)^2 = \mu^2$.  Here $\alpha$, $\beta$,
$\gamma$ and $\delta$ are spin and color indices.  The fields $\overline{d}$ and 
$\overline{f}$ create a down quark and a quark of flavor $f=u$ or $d$ while 
$s$ and $f$ destroy a strange quark and a quark of flavor $f$.  The RI/MOM 
conditions are imposed by removing the four external quark propagators from 
the amplitudes in Eq.~\ref{eq:2010_BK_kinematics}, and then contracting each 
of the resulting seven amputated Green's functions obtained from 
Eq.~\ref{eq:2010_BK_kinematics} with seven projectors 
$\{\Gamma^{ij;f}_{\alpha\beta\gamma\delta}\}_{1 \le j \le 7}$.  The matrix 
$Z^{{\rm lat}\to{\rm RI}}$ is then determined by requiring that the resulting 
49 quantities take their free field values, as is described in detail in 
Refs.~\cite{Blum:2001xb} and \cite{Lehner:2011fz}.

The choice of external momenta specified by Eq.~\ref{eq:2010_BK_kinematics} 
is non-exceptional since no partial sum of these momenta vanish (if their 
signs are chosen so that all four momenta are incoming) and is the choice
used in Refs.~\cite{Li:2008zz} and \cite{Lehner:2011fz}.  Such a choice of 
kinematics is expected to result in normalization conditions which are less 
sensitive to non-zero quark masses and QCD vacuum chiral symmetry breaking 
than would be the case if an exceptional set of momenta had been 
used~\cite{Aoki:2007xm}.  The resulting matrix 
$Z^{{\rm lat}\to{\rm RI}}(\mu,a)/Z_q^2$ obtained for $\mu=2.15$ GeV in 
Ref.~\cite{Li:2008zz} is given in Tab.~\ref{tab:Sam}.  

\begin{table}
\caption{The renormalization matrix $Z^{{\rm lat}\to{\rm RI}}/Z_q^2$ in the seven 
operator chiral basis at the energy scale $\mu=2.15$ GeV.  These values 
were obtained from Ref.~\protect{\cite{Li:2008zz}} by performing an error 
weighted average of the values given in Tabs. 40, 41 and 42 (corresponding
to bare quark masses of 0.01, 0.02 and 0.03) and inverting 
the resulting matrix with an uncorrelated propagation of the errors.  Since
the results given in these three tables are equal within errors, we chose 
to combine them to reduce their statistical errors rather than to perform
a chiral extrapolation.}
  \begin{ruledtabular}
	\begin{tabular}{llllllll}
	  \label{NPRmatrix}
	       & 1         & 2          & 3          & 4          & 5          & 6           & 7 \\
	     1 & 0.825(7) & 0.         & 0.         & 0.         & 0.         & 0.          & 0.\\
	     2 & 0.        & 0.882(38)  & -0.111(41) & -0.009(12)  & 0.010(10) & 0.          & 0.\\
	     3 & 0.        & -0.029(69) & 0.962(92)  & 0.013(22)  & -0.011(25) & 0.          & 0.\\
	     4 & 0.        & -0.04(12)  & -0.01(13)  & 0.924(42)  & -0.149(35) & 0.          & 0.\\
	     5 & 0.        & 0.17(18)   & 0.08(23)   & -0.042(55) & 0.649(63)  & 0.          & 0.\\
	     6 & 0.        & 0.         & 0.         & 0.         & 0.         & 0.943(8)   & -0.154(9)\\
	     7 & 0.        & 0.         & 0.         & 0.         & 0.         & -0.0636(53) & 0.680(11)\\
	\end{tabular}
  \end{ruledtabular}
\label{tab:Sam}
\end{table}

Since these RI/MOM renormalization conditions are being imposed for off-shell,
gauge-fixed external quark lines, we must in principle include a larger number
of operators than the minimal set of seven independent operators which can
represent all gauge invariant matrix elements between physical states of $H_W$.
Therefore, we must also employ a correspondingly larger set of conditions to 
distinguish among this larger set of operators.  This larger set of operators
is required if we are to reproduce with these RI operators all the gauge-fixed, 
off-shell Green's functions that can be constructed using the original, chiral 
basis of lattice operators $Q^\prime_i$.  Thus, as stated in 
Sec.~\ref{sec:DeltaI_3-2}, the relations given in Eq.~\ref{eq:lat2RI} between
the seven lattice and the seven RI operators are valid only when those operators 
appear in physical matrix elements between on-shell states.  For this equation 
to be valid when the operators appear in the off-shell, gauge-fixed Green's that 
define the RI scheme, additional RI/MOM-normalized operators must be added.

However, our ultimate goal is to evaluate on-shell, physical matrix elements 
of these operators.  For such matrix elements there are only seven independent
operators and we can collapse the expanded set of operators referred to above
back to the seven, four-quark, chiral basis operators $Q_i^{\rm RI}$.
This is the meaning of the $7\times7$ matrix $Z^{{\rm lat}\to{\rm RI}}$ matrix 
given in Tab.~\ref{tab:Sam}: gauge symmetry and the equations of motion must 
be imposed to reduce to seven the RI-normalized operators to which the seven 
lattice operators are equated.  In the calculation of $Z^{{\rm lat}\to{\rm RI}}$ 
presented in Ref.~\cite{Li:2008zz} such extra operators are neglected.  For 
all but one, this might be justified because these operators enter only at two 
loops or beyond and the perturbative coefficients that we are using in later 
steps are computed at only one loop.  A single operator, given in Eq.~146 of 
Ref.~\cite{Blum:2001xb} and Eq.~12 of Ref.~\cite{Lehner:2011fz} does appear
at one loop but has also been neglected because it is expected to give a
smaller contribution than other two-quark operators with quadratically 
divergent coefficients whose effects are indeed small.  A final imperfection
in the results presented in Tab.~\ref{tab:Sam} is that the subtraction 
of a third dimension-four, two-quark operator which contains a total derivative 
was not performed.  However, the effect of subtracting this third operator is 
expected to be similar to those of the two operators which were subtracted, 
effects which were not visible outside of the statistical errors 
(see {\it e.g.} Tabs. XIV and XVIII in Ref.~\cite{Blum:2001xb}).

In the second step we convert the seven $RI$ operators obtained above into the 
$\overline{\mathrm{MS}}$ scheme:
\begin{equation}
{Q_i^\prime}^{\overline{\mathrm{MS}}}=\sum_j
               \left(1+\Delta r^{\mathrm{RI}\to\overline{\rm MS}}\right)_{ij}
                Q^\mathrm{RI}_j.
\end{equation}
Here the indices $i$ and $j$ run over the set $\{1,2,3,5,6,7,8\}$ corresponding
to the chiral basis of the operators $Q_j$ defined above and a set of operators
${Q_j^\prime}^{\overline{\mathrm{MS}}}$, with identical chiral properties, which are 
defined in Ref.~\cite{Lehner:2011fz}.  We use the computational framework described 
in Ref.~\cite{Lehner:2011fz} and the resulting $7 \times 7$ matrix 
$\Delta r^{\mathrm{RI}\to\overline{\rm MS}}$ is given in Tab.~VIII of that reference.
As in the case of Eq.~\ref{eq:lat2RI}, the two sets of seven RI and $\overline{\rm MS}$
operators are related by this $7 \times 7$ matrix only when appearing in physical
matrix elements.  Since the values in this table were obtained for the case that 
the wave function renormalization constant for the quark field is the quantity 
$Z^\slashed{q}_q$ it is that factor which we use to extract $Z^{{\rm lat}\to\mathrm{RI}}$ 
from the matrix $Z^{{\rm lat}\to\mathrm{RI}}/Z_q^2$ given in Tab.~\ref{tab:Sam}.  
For our $\beta=2.13$, Iwasaki gauge ensembles $Z^\slashed{q}_q = 0.8016(3)$.
(Note, $Z^\slashed{q}_q$  is the same as the quantity $Z^\prime_q$ introduced
in earlier, exceptional momentum schemes~\cite{Sturm:2009kb}.)

A third and final step is needed before we can combine the Wilson 
coefficients with the matrix elements determined in our calculation to 
obtain the physical amplitudes $A_0$ and $A_2$.  The $7\times7$ matrix 
given in Tab.~VIII of Ref.~\cite{Lehner:2011fz} gives us $\overline{\rm MS}$ 
operators defined in the chiral basis.  However, the Wilson coefficients 
which are available in Ref.~\cite{Buchalla:1995vs} are defined for the 
ten operator basis referred to as basis I in Ref.~\cite{Lehner:2011fz}.  
The conversion between the linearly independent, seven operator basis 
and the conventional set of ten linearly dependent operators is correctly 
given by the application of simple Fierz identities for the case of the 
lattice and RI/MOM operators.  As is explained, for example, in 
Ref.~\cite{Lehner:2011fz}, this procedure is more complex for operators 
defined using $\overline{\rm MS}$ normalization.  Here subtleties of 
defining $\gamma^5$ in dimensions different from four, result in ten 
$\overline{\rm MS}$-normalized operators, $Q_i^{\overline{\rm MS}}$, 
which are not related by the usual Fierz identities, with Fierz 
violating terms appearing at order $\alpha_s$.

Thus, the conventional ten $\overline{\rm MS}$-normalized operators 
$Q_i^{\overline{\rm MS}}$ which appear in Eq.~\ref{eq:weak_eff} must
be constructed, again through one-loop perturbation theory, from 
the seven operators ${Q_i^\prime}^{\overline{\rm MS}}$:
\begin{equation}
Q_i^{\overline{\rm MS}} = \sum_j
               \left(T+\Delta T^{\overline{\rm MS}}_I\right)_{ij} 
               {Q^\prime}^{\overline{\rm MS}}_j,
\end{equation}
in the notation of Ref.~\cite{Lehner:2011fz}.  The $10\times 7$ matrices,
$T$ and $\Delta T^{\overline{\rm MS}}_I$ are given in Eqs.~59 and 65 of
that reference.  (The subscript $I$ on the matrix 
$\Delta T^{\overline{\rm MS}}_I$ identifies the particular ten-operator,
$\overline{\rm MS}$ basis required by the Wilson coefficients of 
Ref.~\cite{Buchalla:1995vs}.)

This entire set of non-perturbative and perturbative transformations
can be summarized by the following equation which expresses the ten
$\overline{\rm MS}$-normalized operators $Q_i^{\overline{\rm MS}}$
in terms of the seven, chiral basis, lattice operators whose matrix
elements we actually compute:
\begin{eqnarray}
Q_i^{\overline{\rm MS}} 
        &=& \sum_j \left[\left(T+\Delta T^{\overline{\rm MS}}_I\right)_{10\times7}
                    \left(1+\Delta r^{{\rm RI}\to\overline{\rm MS}}\right)_{7\times7}
                    \left(Z^{{\rm lat}\to\mathrm{RI}}\right)_{7\times7}\right]_{ij}Q_j^{\rm lat} \\
        &=& \sum_j \left[\left(Z^{{\rm lat}\to \overline{\rm MS}}
                             \right)_{10\times 7}\right]_{ij} Q_j^{\rm lat},
\label{eq:renorm_summary}
\end{eqnarray}
where the subscripts indicate the dimensions of the matrices being
multiplied and the matrix $Z^{{\rm lat}\to \overline{\rm MS}}_{ij}$ is used 
in Eq.~\ref{eq:calAI}.

The physical matrix elements listed in Tabs.~\ref{tab:M_3/2_lat} and 
\ref{Tab:Qi} are obtained by using Eq.~\ref{eq:renorm_summary} to 
determine the matrix elements of the ten conventional operators 
$Q_i^{\overline{\rm MS}}$ in term of the matrix elements of the seven 
lattice operators $Q_j$.  These ten matrix elements are 
then combined with the twenty Wilson coefficients computed for the 
renormalization scale $\mu=2.15$ GeV using the formulae in 
Ref.~\cite{Buchalla:1995vs}.  The values obtained for these Wilson 
coefficients are listed in Tab.~\ref{tab:Wilson_coef}.

\begin{table}
\caption{Wilson Coefficients in the $\overline{MS}$ scheme, at energy 
scale $\mu=2.15$GeV.}
\begin{ruledtabular}
\begin{tabular}{lll}
\label{tab:WilsonCoeff}
$i$ & $y_i^{\overline{MS}}(\mu)$ & $z_i^{\overline{MS}}(\mu)$ \\
\hline
1  &   0	    &	-0.29829     \\
2  &   0            &	1.14439      \\
3  &   0.024141     &	-0.00243827  \\
4  &   -0.058121    &	0.00995157   \\
5  &   0.0102484    &	-0.00110544  \\
6  &   -0.069971    &	0.00657457   \\
7  &   -0.000211182 &	0.0000701587 \\
8  &   0.000779244  &	-0.0000899541\\
9  &   -0.0106787   &	0.0000150176 \\
10 &   0.0029815    &	0.0000656482 \\
\end{tabular}
\end{ruledtabular}
\label{tab:Wilson_coef}
\end{table}

Note, there are many important details of the RI/MOM renormalization 
procedure, such as the subtraction of dimension three and four operators, 
which are not repeated here because they are already discussed with some 
care in Refs.~\cite{Blum:2001xb} and \cite{Lehner:2011fz}.

\bibliography{citations}

\end{document}